\documentclass[pmlr]{jmlr}

\usepackage{booktabs}
\usepackage{array}
\usepackage{longtable}
\usepackage{verbatim}
\usepackage{xcolor}

\jmlrproceedings{}{}
\jmlrvolume{}
\jmlryear{}
\jmlrworkshop{Preprint}
\jmlrpages{}

\newif\ifreview \reviewfalse
\newcommand{\rev}[1]{\ifreview\textcolor{blue}{#1}\else#1\fi}

\newcommand{\R}{\mathbb{R}}
\newcommand{\E}{\mathbb{E}}
\newcommand{\Var}{\mathrm{Var}}
\newcommand{\Cov}{\mathrm{Cov}}
\newcommand{\Pp}{\mathbb{P}}
\newcommand{\toD}{\xrightarrow{\;d\;}}
\newcommand{\toP}{\xrightarrow{\;\Pp\;}}
\newcommand{\op}{o_{\Pp}}
\newcommand{\Op}{O_{\Pp}}
\newcommand{\hth}{\widehat\theta}
\newcommand{\hf}{\widehat f}

\newcommand{\CF}{\mathrm{CF}}
\newcommand{\fstar}{f^\star}
\newcommand{\nPa}[1]{\lVert #1\rVert_{P,\alpha,2}}
\newcommand{\nP}[1]{\lVert #1\rVert_{P,2}}
\makeatletter
\@for\thm@name:=lemma,proposition,corollary,remark\do{%
  \expandafter\let\csname\thm@name\endcsname\relax
  \expandafter\let\csname end\thm@name\endcsname\relax
  \expandafter\let\csname c@\thm@name\endcsname\relax}
\makeatother
\newtheorem{lemma}{Lemma}
\newtheorem{proposition}{Proposition}
\newtheorem{corollary}{Corollary}
\theorembodyfont{\upshape}
\newtheorem{remark}{Remark}
\theorembodyfont{\itshape}
\newtheorem{assumption}{Assumption}
\newenvironment{proofof}[1]{\par\medskip\noindent\textbf{Proof of #1.}\enspace\ignorespaces}{\hfill$\blacksquare$\par\medskip}
\makeatletter\let\ExpInput\@@input\makeatother
\title[Design-Based Inference with AI Predictions]{Design-Based Inference with AI Predictions}

\author{%
 \Name{Victor Chernozhukov\nametag{$^{*,\dagger}$}},
 \Name{Jiaxuan Li\nametag{$^{*,\ddagger}$}},
 \Name{Anqi Ni\nametag{$^{*,\S}$}},
 \Name{Ali Horta\c{c}su\nametag{$^{*,\P}$}},
 \Name{Bohan Zhang\nametag{$^{*,\|}$}}
}

\begin{document}
\maketitle
{\renewcommand{\thefootnote}{\fnsymbol{footnote}}%
\footnotetext[1]{All authors contributed equally. The authors used AI interactively to develop their ideas, derive key supporting results, and assist with verification and writing. They take full responsibility for the paper's content.}%
\footnotetext[2]{Massachusetts Institute of Technology; vchern@mit.edu.}%
\footnotetext[3]{Independent Researcher; lijiaxuan0529@gmail.com.}%
\footnotetext[4]{University of Chicago; anqini4@gmail.com.}%
\footnotetext[5]{University of Chicago; hortacsu@uchicago.edu.}%
\footnotetext[6]{University of Michigan; zbohan@umich.edu.}}

\begin{abstract}
We extend and refine prediction-powered inference for linear functionals of a regression, including treatment effects and average derivatives, when outcome labels are scarce. The study design supplies a known Riesz representer, whose weights give an exact correction for prediction errors. Under independent labeled and target sampling, the resulting estimator is unbiased for every admissible fixed predictor. We fit AI-assisted predictors to minimize the empirical variance of weighted residuals. We establish asymptotic variance optimality within a possibly misspecified prediction class when the labeled sample is asymptotically smaller than the target sample. Linear prediction classes yield transformed least squares. Our conditions for valid cross-fitted inference allow a misspecified limit without imposing a regression convergence rate. We also develop full-variance optimization through direct fitting and stacking at general sample-size ratios. Resampling the OpenResearch Unconditional Income Study illustrates these precision gains. With 10\% of each arm labeled, stacking learners that combine few-shot predictions, covariates, and AI-generated features reduces variance relative to labeled-only estimation by 15.3--45.6\% across fourteen outcomes. Its 95\% intervals contain the full-cohort benchmark in 94.7\% of replicates on average.
\end{abstract}

\section{Introduction}
\label{sec:intro}

We develop a framework for using AI predictions to improve inference on a broad class of population parameters when outcome labels are scarce. The framework covers linear functionals of a regression, including means, treatment effects, average derivatives, and policy counterfactuals. Table~\ref{tab:examples} illustrates the range of parameters and the common structure that connects them. Our central requirement is that the study design or a specified model supply the parameter's \emph{Riesz representer}: the weighting function that translates regression errors into errors in the target parameter. We exploit this knowledge both to correct prediction errors exactly and to fit predictors for inferential precision.

Knowledge of representers arises naturally in several settings. In randomized controlled trials, the randomization protocol specifies treatment-assignment probabilities. In algorithmic quasi-experiments, a known allocation mechanism can likewise supply the relevant probabilities, for example through lotteries or randomized tie-breaking. In structural counterfactual analysis, the model may specify the reference and counterfactual distributions, making the density ratios or scores needed for the analysis known.

To describe the framework, let $P$ be the distribution of labeled observations $(Y,X)$, and let $\gamma_0(x):=\E_P[Y\mid X=x]$. Target observations $Z$ have distribution $Q$. A known map $m(z,\cdot)$, linear in its second argument, specifies the parameter
\begin{equation}
\label{eq:theta_def}
\theta_0:=\E_Q\,m(Z,\gamma_0).
\end{equation}
For example, in a randomized experiment, $X=(D,W)$ contains treatment $D\in\{0,1\}$ and baseline covariates $W$. Taking $Z=W$ and $m(W,\gamma_0)=\gamma_0(1,W)-\gamma_0(0,W)$ gives the average treatment effect.

Replacing $\gamma_0$ by an AI-assisted predictor $f$ generally changes the target. The known representer $\alpha_0$ supplies the correction. For a fixed $f$, the conditional mean of $Y-f(X)$ given $X$ is $\gamma_0(X)-f(X)$, so weighted outcome residuals have the same expectation as weighted regression errors. The defining property of the representer converts this expectation into the discrepancy in the target. Under the conditions of Section~\ref{sec:framework}, every admissible predictor $f$ therefore satisfies
\begin{equation}
\label{eq:intro_identity}
\theta_0 \;=\; \E_Q\, m(Z,f) \;+\; \E_P\, \alpha_0(X)\{Y-f(X)\}.
\end{equation}
The first term evaluates the functional using predictions for target observations, and the second corrects its error using weighted residuals on labeled observations. Under the conditions of Section~\ref{sec:framework}, replacing the expectations in \eqref{eq:intro_identity} by sample averages gives an exactly unbiased estimator for every admissible fixed predictor. Prediction quality affects its variance through the target term and the weighted residuals.

Because the representer is known, the correction remains exact even for a misspecified predictor. Under independent sampling and integrability, cross-fitting preserves this identity after fitting: conditional on the training observations, the evaluation observations retain their original distributions, and \eqref{eq:intro_identity} applies to the realized predictor. Under the stated moment and continuity conditions, inference remains valid when the fitted predictors converge to a deterministic limit, without imposing a regression convergence rate.

This freedom to use misspecified predictors makes inferential precision a natural fitting objective. \emph{Fine-tuning} fits a prediction rule from fixed AI outputs and covariates to labeled outcomes. We also develop procedures that fit directly for the estimator's variance. Let $T$ and $N$ denote the labeled and target sample sizes. When $T=o(N)$, the labeled correction determines the leading variance, suggesting minimization of the variance of its weighted residual. For a linear prediction class, this is ordinary least squares with a free intercept after multiplying both the outcome and the regressors by the known representer. Under the stated conditions, the fit asymptotically attains the smallest leading variance within the chosen class. We also develop full-variance optimization through direct fitting and stacking without requiring $T=o(N)$ (Section~\ref{sec:construction} and Appendix~\ref{app:full_variance_fit}).

The framework connects design-based weighting and regression adjustment \citep{horvitz1952ht,robins1994estimation,lin2013agnostic}, orthogonal estimation and Riesz representations \citep{chernozhukov2018dml,chernozhukov2022riesz,chernozhukov2025covshift}, and prediction-powered inference \citep{angelopoulos2023science,angelopoulos2024ppiplusplus,zrnic2024cross}. For a mean without distribution shift, our correction recovers PPI, and PPI++ already optimizes a scalar adjustment for variance. Our focus is the common known-representer structure across the functionals in Table~\ref{tab:examples} and valid inference after fitting possibly misspecified prediction classes for precision. Appendix~\ref{app:related} develops these connections and discusses calibration, prognostic adjustment, and related methods.

We examine the precision gains in the OpenResearch Unconditional Income Study (ORUS), a randomized trial that paid 1,000 adults \$1,000 per month for three years \citep{vivalt2024,bartik2024}. A language model predicts each participant's outcome under the \$1,000 and \$50 payments. We draw labeled and target samples independently with replacement within each arm, with a 10\% label budget, and estimate fourteen effects. Relative to labeled-only estimation, the bias-corrected estimator using a fixed prediction reduces variance by 8.5--42.7\%. Stacking learners that combine few-shot predictions, covariates, and AI-generated features yields reductions of 15.3--45.6\%. Its 95\% intervals contain the full-cohort difference in means in 94.7\% of replicates on average. Prediction-loss, residual-variance, and full-variance fitting yield similar precision, with a slight average advantage for prediction loss (Appendix~\ref{app:full}). The source records and assignments remain fixed, so coverage concerns the difference between the two empirical arm means.

In what follows, Sections~\ref{sec:framework} and~\ref{sec:finetune} establish the identity and cross-fitting results that support the learning procedure. Section~\ref{sec:construction} constructs predictors from AI outputs, covariates, and the representer, establishes inference after prediction-loss fitting, and develops variance optimization and stacking. Section~\ref{sec:application} examines precision gains from AI predictions and calibration by resampling the OpenResearch Unconditional Income Study. The appendix contains proofs and additional theoretical and empirical results.

\section{Framework: Known Design and an Unbiased Estimator}
\label{sec:framework}

\subsection{Setup and the key identity: Linear Functionals and Known Representers}

We take a superpopulation perspective: the observations are draws from distributions, and the target is a functional of those distributions. We use a labeled sample to correct prediction errors and an unlabeled target sample to evaluate the functional. The labeled observations $\{(Y_t,X_t)\}_{t=1}^{T}$ are drawn from a distribution $P$ on $\R\times\mathcal X$, and the target observations $\{Z_i\}_{i=1}^{N}$ are drawn from a distribution $Q$ on $\mathcal Z$. Each sample consists of independent and identically distributed observations, and the two samples are independent. Although labels are often scarce, we impose no restriction on the relative sizes of $N$ and $T$.

Assume $\E_PY^2<\infty$ and let $\gamma_0(x):=\E_P[Y\mid X=x]$ denote the regression function. Write $\nP{f}:=(\E_P f(X)^2)^{1/2}$, and work on a linear space $\mathcal L$ of square-integrable functions containing $\gamma_0$. The known map $m:\mathcal Z\times\mathcal L\to\R$ specifies the target parameter in \eqref{eq:theta_def}.
For example, to estimate the population mean, we take $Z=X$, let $Q$ be the marginal distribution of $X$ under $P$, and set $m(z,f):=f(z)$. Then $\theta_0=\E_P\gamma_0(X)=\E_PY$. For other functionals, $m$ may take a contrast or a derivative of the regression, as illustrated in Table~\ref{tab:examples}; see \citet{chernozhukov2022autodml} for further examples.

\begin{table}[!b]
\centering
\caption{Examples of linear functionals with known Riesz representers. Here $W$ denotes baseline covariates, $D$ a treatment or dose, and $P_X$ the marginal distribution of $X$ under $P$. In Examples 1--4, $Q$ is the distribution of $Z$ induced by $P$; in Example 5, $Q$ is a target covariate distribution that may differ from $P_X$.}
\label{tab:examples}
\footnotesize\setlength{\tabcolsep}{3pt}
\renewcommand{\arraystretch}{1.05}
\begin{tabular}{@{}r@{\hspace{0.55em}}>{\raggedright\arraybackslash}p{0.18\linewidth}>{\raggedright\arraybackslash}p{0.32\linewidth}>{\raggedright\arraybackslash}p{0.24\linewidth}>{\raggedright\arraybackslash}p{0.18\linewidth}@{}}
\toprule
\multicolumn{2}{@{}l}{Example} & Design knowledge & $m(z,f)$ & $\alpha_0(x)$ \\
\midrule
1. & Mean without\newline covariate shift\textsuperscript{a} & labels sampled at random; $Z=X$ & $f(z)$ & $1$ \\
\addlinespace[6pt]
2. & Effect in an RCT & known propensity $\pi(w):=P(D=1\mid W=w)$ with $0<\pi(w)<1$; $X=(D,W)$, $Z=W$ & $f(1,w)-f(0,w)$ & $\dfrac{d}{\pi(w)}-\dfrac{1-d}{1-\pi(w)}$ \\
\addlinespace[6pt]
3. & Average derivative\textsuperscript{b} & known density $\varrho(d\mid w)$; $Z=X=(D,W)$ & $\partial_d f(d,w)$ & $-\partial_d\log\varrho(d\mid w)$ \\
\addlinespace[6pt]
4. & Dose shift\newline $D\mapsto D+\delta$\textsuperscript{c} & known conditional density or pmf $\varrho(d\mid w)$; fixed $\delta$; $Z=X=(D,W)$ & $f(d+\delta,w)-f(d,w)$ & $\dfrac{\varrho(d-\delta\mid w)}{\varrho(d\mid w)}-1$ \\
\addlinespace[6pt]
5. & Mean under\newline covariate shift\textsuperscript{d} & known density ratio $r:=dQ/dP_X$ & $f(z)$ & $r(x)$ \\
\bottomrule
\end{tabular}
\par\smallskip
\begin{minipage}{\linewidth}
\footnotesize\raggedright
\textsuperscript{a}\,Example 1 recovers prediction-powered inference (PPI) for a mean; \textsuperscript{b}\,Average derivatives require additional smoothness, boundary, and moment conditions; see Appendix~\ref{app:derivative_regularity}; \textsuperscript{c}\,For almost every $w$, the conditional distribution of $D+\delta$ given $W=w$ must be absolutely continuous with respect to that of $D$ given $W=w$. The density or pmf is taken as zero outside its support, and the displayed representer must be square-integrable under $P$; \textsuperscript{d}\,We require $Q\ll P_X$ and $\E_P r(X)^2<\infty$. Interpreting the parameter as the target population's outcome mean requires that the conditional mean of the outcome remain $\gamma_0$.
\end{minipage}
\end{table}

Replacing $\gamma_0$ by a predictor $f$ generally changes the target functional. To correct this discrepancy using labeled observations, we need to express the effect of a regression error as a weighted average under $P$. As explained in the Introduction, knowledge of Riesz representers arises naturally in randomized controlled trials, algorithmic quasi-experiments, and structural counterfactual analysis.

When the relevant probabilities or distributions are known, they determine the Riesz representer for the chosen target, subject to the support and moment conditions illustrated in Table~\ref{tab:examples}. The outcome regression can remain unknown and difficult to estimate. This separation between knowledge of the assignment or distributional mechanism and knowledge of the outcome response is what permits an exact correction for prediction errors. The following assumption states the required representation.

\begin{assumption}[Linearity and a known representer]
\label{ass:known_riesz}
For each $z$, the map $m(z,\cdot)$ is linear on $\mathcal L$, and $m(Z,f)$ is measurable and integrable under $Q$ for every $f\in\mathcal L$. There is a \emph{known} measurable function $\alpha_0:\mathcal X\to\R$ with $\E_P\,\alpha_0(X)^2<\infty$ such that
\begin{equation}
\label{eq:representer}
\E_Q\, m(Z,\Delta)\;=\;\E_P\,\alpha_0(X)\Delta(X)
\qquad\text{for all }\Delta\in\mathcal L.
\end{equation}
\end{assumption}

The function $\alpha_0$ is a \emph{Riesz representer}: equation~\eqref{eq:representer} expresses the linear functional $\Delta\mapsto\E_Qm(Z,\Delta)$ as an inner product under $P$ \citep{newey1994asymptotic}. Its square-integrability also makes this functional continuous in the norm $\nP{\cdot}$, by Cauchy--Schwarz. The assignment or sampling rule supplies these weights. Estimating the representer generally leaves a bias involving its error and the prediction error $\gamma_0-f$, so exact correction for an arbitrary predictor no longer follows. This case requires additional error control, as in \citet{chernozhukov2022riesz,chernozhukov2025covshift}.

To correct the error from using a predictor $f$, we apply the representer identity to the regression difference $\gamma_0-f$. The following lemma expresses the resulting correction in terms of the observable residual $Y-f(X)$.

\begin{lemma}[Unbiasedness identity]
\label{lem:identity}
Under Assumption~\ref{ass:known_riesz}, every $f\in\mathcal L$ satisfies
\begin{equation}
\label{eq:identity}
\theta_0=\E_Q\, m(Z,f)+\E_P\,\alpha_0(X)\{Y-f(X)\}.
\end{equation}
\end{lemma}

To see why the correction works, we first condition on $X$ and then use linearity and \eqref{eq:representer} with $\gamma_0-f\in\mathcal L$ to obtain
\[
\E_P\,\alpha_0(X)\{Y-f(X)\}
=\E_P\,\alpha_0(X)\{\gamma_0(X)-f(X)\}
=\theta_0-\E_Q\,m(Z,f).
\]

Thus the weighted residual cancels the discrepancy exactly, regardless of the predictor's accuracy. For a fixed predictor, we obtain an unbiased estimator by replacing the two expectations with sample averages, as we do next.

\subsection{The estimator with a fixed predictor}

Let $f\in\mathcal L$ be fixed, that is, chosen without reference to either sample; a pretrained language model is a leading example. Replacing the two expectations in \eqref{eq:identity} by sample averages gives the \emph{known-Riesz estimator}
\begin{equation}
\label{eq:estimator}
\hth(f):=\frac1N\sum_{i=1}^{N}m(Z_i,f)+\frac1T\sum_{t=1}^{T}\alpha_0(X_t)\{Y_t-f(X_t)\}.
\end{equation}
The target contribution has variance $\sigma_Q^2(f):=\Var_Q\,m(Z,f)$, and the weighted residual has variance $\sigma_P^2(f):=\Var_P\bigl(\alpha_0(X)\{Y-f(X)\}\bigr)$. The estimator's variance and its sample analogue are
\[
\sigma^2(f):=\frac{\sigma_Q^2(f)}{N}+\frac{\sigma_P^2(f)}{T},
\qquad
\widehat V(f):=\frac{\widehat\sigma_Q^2(f)}{N}+\frac{\widehat\sigma_P^2(f)}{T}.
\]
Here $\widehat\sigma_Q^2(f)$ and $\widehat\sigma_P^2(f)$ are the sample variances of the summands in \eqref{eq:estimator}, and the dependence of $\sigma^2(f)$ on $N$ and $T$ is left implicit. The next lemma combines the unbiasedness identity with independence of the samples.

\begin{lemma}[Unbiasedness and variance]
\label{lem:unbiased_var}
Let Assumption~\ref{ass:known_riesz} hold, and let $f\in\mathcal L$ be fixed. Then $\E\,\hth(f)=\theta_0$. If additionally $\sigma_Q^2(f)<\infty$ and $\sigma_P^2(f)<\infty$, then $\Var\,\hth(f)=\sigma^2(f)$.
\end{lemma}

Taking expectations in \eqref{eq:estimator} gives unbiasedness by Lemma~\ref{lem:identity}. This step uses only the marginal laws. Independence within and between the samples gives the variance formula and, with finite summand variances, the following normal approximation.

\begin{theorem}
\label{thm:clt}
Let Assumption~\ref{ass:known_riesz} hold, and let $f\in\mathcal L$ be fixed with $\sigma_Q^2(f)<\infty$, $\sigma_P^2(f)<\infty$, and $\sigma^2(f)>0$. If $N\to\infty$ and $T\to\infty$ in any manner whatever, then
\[
\frac{\hth(f)-\theta_0}{\sigma(f)}\toD\mathcal N(0,1)
\qquad\text{and}\qquad
\frac{\hth(f)-\theta_0}{\widehat V(f)^{1/2}}\toD\mathcal N(0,1).
\]
\end{theorem}

The correction ensures validity even when $f$ is inaccurate; precision depends on both variance components. Theorem~\ref{thm:clt} imposes no restriction on $N/T$. When both components are positive and $N/T\to\xi\in[0,\infty)$, we have $\sigma^2(f)\sim\{\sigma_Q^2(f)+\xi\sigma_P^2(f)\}/N$. When $T=o(N)$, the labeled correction determines the leading variance, $\sigma^2(f)\sim\sigma_P^2(f)/T$. The next proposition shows why the true regression is optimal for this component.

\begin{proposition}
\label{prop:oracle}
Suppose that $\E_P\,\alpha_0(X)^2Y^2<\infty$. Among all $f\in\mathcal L$ with $\sigma_P^2(f)<\infty$, the choice $f=\gamma_0$ minimizes $\sigma_P^2(f)$, and
$\sigma_P^2(f)=\sigma_P^2(\gamma_0)+\Var_P\bigl(\alpha_0(X)\{\gamma_0(X)-f(X)\}\bigr)$.
\end{proposition}

To see the decomposition, write the weighted residual as the sum of $\alpha_0(X)\{Y-\gamma_0(X)\}$ and $\alpha_0(X)\{\gamma_0(X)-f(X)\}$. The first term has conditional mean zero given $X$, so it is uncorrelated with the second. The additional variance is nonnegative and vanishes at $f=\gamma_0$, though other predictors can also minimize it. Along $f_t:=(1-t)f+t\gamma_0$, $0\le t\le1$, this excess variance shrinks by $(1-t)^2$; validity requires no such contraction.

For general sample-size ratios, we can optimize the full variance $\sigma^2(f)$ over a chosen prediction class, accounting also for the target-sample contribution $\sigma_Q^2(f)/N$, where $\sigma_Q^2(f)=\Var_Qm(Z,f)$. Appendix~\ref{app:full_variance_fit} develops this approach. In our experiments, full-variance fitting modestly improves average precision over residual-variance fitting (Appendix~\ref{app:full}). We focus on the simpler residual-variance criterion in the main text to convey the central ideas.

\section{Fine-tuning with cross-fitting}
\label{sec:finetune}

We now allow the predictor to be fitted to the labeled observations. Reusing those observations in the correction term can invalidate the unbiasedness argument. For example, if $\hf$ interpolates the labels, every fitted residual is zero, and the estimator reduces to the target average of the fitted predictions. Cross-fitting preserves the correction by evaluating each predictor on observations excluded from its training sample \citep{schick1986efficient}. Appendices~\ref{app:insample_validity}--\ref{app:insample_stability} study inference without cross-fitting and give conditions for asymptotically valid inference based on control of the prediction class's complexity or the fitting algorithm's stability.

Fix a partition of $\{1,\dots,T\}$ into $K$ folds $I_1,\dots,I_K$ with weights $\omega_k:=|I_k|/T\ge\underline\omega>0$, where $K$ and $\underline\omega$ do not change with $T$. Each $\hf_k$ is fitted from the labeled observations outside $I_k$ and, if desired, external data independent of both samples, but not from the target sample. Section~\ref{sec:construction} gives concrete constructions of these predictors using AI outputs, covariates, and the known representer. We require $\hf_k\in\mathcal L$ almost surely and all evaluations below to be measurable. The \emph{cross-fitted known-Riesz estimator} is
\begin{equation}
\label{eq:theta_cf}
\hth_{\CF}:=\sum_{k=1}^{K}\omega_k\,\hth_k,
\qquad
\hth_k:=\frac1N\sum_{i=1}^{N}m(Z_i,\hf_k)+\frac1{|I_k|}\sum_{t\in I_k}\alpha_0(X_t)\{Y_t-\hf_k(X_t)\}.
\end{equation}

\begin{proposition}[Exact unbiasedness under cross-fitting]
\label{prop:cf_unbiased}
Under Assumption~\ref{ass:known_riesz}, suppose that the fits constructed above satisfy $\hf_k\in\mathcal L$ almost surely, $\E\,\sigma_Q^2(\hf_k)<\infty$, and $\E\,\sigma_P^2(\hf_k)<\infty$ for each $k$. Then $\E\,\hth_{\CF}=\theta_0$, whatever algorithm produces these fits.
\end{proposition}

For each $k$, conditioning on $\hf_k$ leaves the observations in $I_k$ and the target sample with their original independent laws. Lemma~\ref{lem:identity} therefore gives $\E[\hth_k\mid\hf_k]=\theta_0$. The expected variance conditions ensure that $\hth_k$ is integrable, so we may average over the fit and then over folds. The fold estimates need not be independent for this argument.

Exact unbiasedness alone does not give a normal approximation. For inference, we ask that the fits stabilize at a deterministic limit. To measure their error in both the prediction and the weighted correction, define $\nPa{f}:=\bigl(\E_P\{1+\alpha_0(X)^2\}f(X)^2\bigr)^{1/2}$.

\begin{assumption}[Pseudo-consistency]
\label{ass:pseudo}
For some deterministic function $\fstar\in\mathcal L$ with $\sigma_Q^2(\fstar)$ and $\sigma_P^2(\fstar)$ finite, $\nPa{\hf_k-\fstar}=\op(1)$ for each $k\le K$.
\end{assumption}

The limit $\fstar$ may be misspecified, and no convergence rate is imposed. When $\alpha_0$ is bounded, the weighted norm is equivalent to $\nP{\cdot}$. We also need convergence of predictions to carry over to their contributions to the target functional.

\begin{assumption}[Mean-square continuity]
\label{ass:msc}
There are a fixed class $\mathcal D\subseteq\mathcal L$ and constants $M<\infty$ and $a\in(0,2]$ such that $\E_Q\, m(Z,\Delta)^2\le M\nP{\Delta}^{a}$ for every $\Delta\in\mathcal D$, and $\Pp(\hf_k-\fstar\in\mathcal D\text{ for all }k\le K)\to1$.
\end{assumption}

Assumption~\ref{ass:msc} holds with $a=2$ and $\mathcal D=\mathcal L$ for the mean without covariate shift, and for the treatment effect when the propensity is bounded away from 0 and 1. Under weaker overlap, a square-integrable representer and uniformly bounded deviations give $a=1$ for the treatment effect. Appendix~\ref{app:verify} gives the verifications for all five examples, including conditions for derivatives. The next lemma shows how these assumptions control the effect of fitting.

\begin{lemma}
\label{lem:quasi_oracle}
Let $\Delta_k:=\hf_k-\fstar$. Under Assumptions~\ref{ass:known_riesz}--\ref{ass:msc},
\[
\hth_{\CF}-\hth(\fstar)
=\Op\Bigl(N^{-1/2}\max_{k\le K}\nP{\Delta_k}^{a/2}+T^{-1/2}\max_{k\le K}\nPa{\Delta_k}\Bigr).
\]
\end{lemma}

The representer identity cancels the population prediction errors, leaving centered sample averages. Their fluctuations are controlled by the two norms above. Under mild conditions, this remainder is negligible relative to the standard error.

For inference, we estimate the variance using
\begin{equation}
\label{eq:Vcf}
\widehat V_{\CF}:=\frac1N\,\widehat{\Var}_N\!\left(\sum_{k=1}^{K}\omega_k\,m(Z_i,\hf_k)\right)+\frac1T\,\widehat{\Var}_T\!\left(\alpha_0(X_t)\{Y_t-\hf_{k(t)}(X_t)\}\right),
\end{equation}
where $k(t)$ is the fold containing $t$, and $\widehat\Var_N$ and $\widehat\Var_T$ are sample variances over the target and labeled observations. All fits use the same target sample, so we average their target contributions before taking the sample variance. Each labeled residual instead uses its own held-out fit.

\begin{theorem}
\label{thm:cf_clt}
Let Assumptions~\ref{ass:known_riesz}--\ref{ass:msc} hold with $\sigma_Q^2(\fstar)>0$ and $\sigma_P^2(\fstar)>0$, and let $N\to\infty$ and $T\to\infty$ in any manner whatever. Then
\[
\frac{\hth_{\CF}-\theta_0}{\sigma(\fstar)}\toD\mathcal N(0,1)
\qquad\text{and}\qquad
\frac{\widehat V_{\CF}}{\sigma^2(\fstar)}\toP 1.
\]
\end{theorem}

Hence $\hth_{\CF}\pm1.96\,\widehat V_{\CF}^{1/2}$ is an asymptotically valid 95\% confidence interval. The cross-fitted estimator has the same first-order behavior as if $\fstar$ were fixed in advance. Thus, fine-tuning can improve precision without a first-order cost from estimating the predictor, even when its limit is misspecified. Appendix~\ref{app:proof_crossfitting} gives the proofs.

\section{Building the predictor from AI outputs}
\label{sec:construction}

We now construct the predictors $\hf_k$ used in Section~\ref{sec:finetune}. Our basic approach predicts the labeled outcomes using AI outputs, observed covariates, and the known representer, fitting separately outside each fold. We also consider fitting criteria that target the estimator's variance directly.

\subsection{Prediction classes and the representer}
\label{sec:prediction_classes}

Let $p:\mathcal X\to\R^{d_p}$ be the raw output of an AI system: one prediction, the answers to several prompts, or an embedding. We do not retrain the system. We choose a low-dimensional map from its output, the representer, and a block $h(x)$ of covariates, forming the admissible family
\begin{equation}
\label{eq:template}
f_\beta(x):=F\bigl(p(x),\alpha_0(x),h(x);\beta\bigr),\qquad
\mathcal F:=\{f_\beta:\beta\in\mathcal B\}\subseteq\mathcal L.
\end{equation}
We do not require $\gamma_0\in\mathcal F$.

For a scalar prediction $p$, useful choices include (a)~\emph{no fit}, $f=p$, recovering PPI for a mean without covariate shift; (b)~\emph{one scale parameter}, $f=\lambda p$; (c)~\emph{calibration}, $f=b+c\,p+v\,\alpha_0$, or in an experiment $f(d,w)=b_d+c\,p(d,w)$; (d)~\emph{augmented calibration}, adding $\beta^\top h$ to the preceding specifications, where $h$ may contain covariates, further AI outputs, or AI-generated features; (e)~\emph{shape-restricted calibration}, $f=\iota(p)+v\,\alpha_0$ with $\iota$ nondecreasing; and (f)~\emph{nonlinear maps} of $(p,\alpha_0,h)$, such as boosted trees or a small network. These choices specify the prediction class; the fitting criteria below determine how it is trained.

The representer is a natural input because the conditional mean of $Y$ given $\alpha_0(X)$ determines the target. Define $A:=\alpha_0(X)$.

\begin{lemma}
\label{lem:index}
Let Assumption~\ref{ass:known_riesz} hold, and let $\bar\gamma_0(a):=\E_P[Y\mid A=a]$. Then $\theta_0=\E_P\,A\,\bar\gamma_0(A)$. If $\bar\gamma_0\circ\alpha_0\in\mathcal L$, this also equals $\E_Q\,m(Z,\bar\gamma_0\circ\alpha_0)$.
\end{lemma}

The lemma restates Lemma~3.1 of \citet{RieszNetForestRiesz} and generalizes the propensity-score reduction of \citet{rosenbaum1983central}; its proof is in Appendix~\ref{app:prediction_classes}. For the average treatment effect in an experiment with a constant propensity, $\alpha_0$ takes one value per arm, so a free function of the representer is a pair of arm intercepts.

Including the representer as a regressor also has a direct least-squares interpretation. If a population squared-loss minimizer $\fstar$ satisfies $\fstar+t\alpha_0\in\mathcal F$ for all sufficiently small positive and negative $t$, the loss has zero derivative at $t=0$, giving $\E_P[\alpha_0(X)\{Y-\fstar(X)\}]=0$. The correction term in Lemma~\ref{lem:identity} therefore vanishes, yielding $\E_Qm(Z,\fstar)=\theta_0$, even when $\fstar$ is misspecified.

\subsection{Fitting by prediction loss}
\label{sec:prediction_loss}

The simplest approach fits the predictor to the observed outcomes. On the labeled observations outside fold $I_k$, let $n_k:=T-|I_k|$ and compute
\begin{equation}
\label{eq:prediction_loss_fit}
\widehat\beta_k\in\arg\min_{\beta\in\mathcal B}
\left\{\frac1{n_k}\sum_{t\notin I_k}\{Y_t-f_\beta(X_t)\}^{2}+\lambda_k\Omega(\beta)\right\},
\qquad \hf_k:=f_{\widehat\beta_k},
\end{equation}
where $\Omega$ is a penalty and $\lambda_k\ge0$ controls regularization. Taking $\lambda_k=0$ gives unpenalized least squares within the chosen family; ridge and lasso use quadratic and absolute-value penalties in a linear family.

The population target is the best squared-error predictor in the family. It need not equal the regression function: Section~\ref{sec:finetune} requires convergence to a fixed predictor in the weighted norm of Assumption~\ref{ass:pseudo}. For compact finite-dimensional families, the following result gives this weighted convergence.

\begin{proposition}[Prediction-loss fitting]
\label{prop:param}
Let $\mathcal B\subseteq\R^d$ be nonempty and compact, with $d$ fixed, and let $f_\beta\in\mathcal L$. Suppose that $\beta\mapsto f_\beta(X)$ is almost surely continuous, $\Omega$ is continuous, and $\E_P\sup_{\beta\in\mathcal B}\{1+\alpha_0(X)^2\}f_\beta(X)^2<\infty$. Suppose also that $\lambda_k\toP0$ and that every minimizer of $\E_P\{Y-f_\beta(X)\}^2$ represents the same function $\fstar$, $P$-almost surely. Then, for each $k$, every measurable fit in \eqref{eq:prediction_loss_fit} satisfies $\nPa{\hf_k-\fstar}=\op(1)$ as $n_k\to\infty$.
Under the remaining conditions of Theorem~\ref{thm:cf_clt}, the resulting cross-fitted estimator satisfies its inferential conclusions.
\end{proposition}

The result requires neither correct specification nor a particular convergence rate. Appendix~\ref{app:prediction_loss} gives the proof and critical-radius bounds that characterize convergence rates for bounded least-squares classes, including weighted-norm rates when the representer is bounded.

\subsection{Fitting for estimator variance}
\label{sec:variance_fitting}

Prediction-loss fitting need not minimize the estimator's variance, so we now fit the predictor to minimize variance, at least to leading order. When $T=o(N)$, this amounts to minimizing $\sigma_P^2(f)$, the variance of the weighted residual $\alpha_0(X)\{Y-f(X)\}$. We retain both variance components in the standard errors.

For each fold, we estimate this variance using the labeled observations outside $I_k$. Write $R_f:=\alpha_0(X)\{Y-f(X)\}$ and let $\mathbb P_{-k}g:=n_k^{-1}\sum_{t\notin I_k}g(Y_t,X_t)$ denote their empirical average. The empirical residual variance is $\widehat\sigma_{P,k}^2(f):=\mathbb P_{-k}R_f^2-(\mathbb P_{-k}R_f)^2$.

To minimize this criterion by least squares, introduce a centering constant $c\in\R$ and consider $\mathbb P_{-k}(R_f-c)^2$. For fixed $f$, minimizing over $c$ gives $c=\mathbb P_{-k}R_f$ and minimum $\widehat\sigma_{P,k}^2(f)$. Thus we compute
\begin{equation}
\label{eq:variance_fit}
(\widehat\beta_k,\widehat c_k)\in\arg\min_{\beta\in\mathcal B,\,c\in\R}
\frac1{n_k}\sum_{t\notin I_k}\bigl[\alpha_0(X_t)\{Y_t-f_\beta(X_t)\}-c\bigr]^2.
\end{equation}
For a linear family $f_\beta(x)=q(x)^\top\beta$, this is least squares of $\alpha_0Y$ on $\alpha_0q(X)$ with an intercept, subject to $\beta\in\mathcal B$. We use $\hf_k:=f_{\widehat\beta_k}$ in the cross-fitted estimator \eqref{eq:theta_cf}, with variance estimate \eqref{eq:Vcf}. The fitted $\widehat c_k$ centers only the training criterion; the correction in \eqref{eq:theta_cf} uses the uncentered residual $R_{\hf_k}$.

The next result bounds the excess population variance by the errors in estimating and minimizing the empirical criterion.

\begin{proposition}[Residual-variance optimization and inference]
\label{prop:variance_fit}
Let $\mathcal F\subseteq\mathcal L$ be fixed and nonempty, with $\E_PR_f^2<\infty$ for every $f\in\mathcal F$. For each $k$, let $\hf_k\in\mathcal F$ be a measurable training-fold fit satisfying $\widehat\sigma_{P,k}^2(\hf_k)\le\inf_{f\in\mathcal F}\widehat\sigma_{P,k}^2(f)+\eta_k$, with $\eta_k\ge0$, and let $\varepsilon_k:=\sup_{f\in\mathcal F}|\widehat\sigma_{P,k}^2(f)-\sigma_P^2(f)|$ be measurable. Then
\begin{equation}
\label{eq:variance_oracle}
0\le\sigma_P^2(\hf_k)-\inf_{f\in\mathcal F}\sigma_P^2(f)\le2\varepsilon_k+\eta_k.
\end{equation}
If $\varepsilon_k+\eta_k\toP0$ for every $k$, the fits are variance-optimal in probability. If additionally Assumptions~\ref{ass:known_riesz}--\ref{ass:msc} hold with $\sigma_P^2(\fstar)>0$ and $\sigma_Q^2(\fstar)>0$, then $\sigma_P^2(\fstar)=\inf_{f\in\mathcal F}\sigma_P^2(f)$. When also $T=o(N)$ as $N,T\to\infty$, we have $\sigma^2(\fstar)\sim T^{-1}\inf_{f\in\mathcal F}\sigma_P^2(f)$. Under the remaining conditions of Theorem~\ref{thm:cf_clt}, the resulting cross-fitted estimator satisfies its inferential conclusions.
\end{proposition}

Appendix~\ref{app:variance_fit} gives the proof and sufficient conditions for uniform convergence of the empirical criterion and weighted convergence of the fits. A normalization may be needed to identify the limiting predictor: for a mean, adding a constant to $f$ leaves $\sigma_P^2(f)$ unchanged. The result permits misspecification and requires no rate of convergence to $\fstar$.

Full-variance optimization can select a different predictor when $T/N$ does not vanish. For a mean without covariate shift, the coefficient minimizing the full variance within $f=\lambda p$ is $\lambda_{\mathrm{LS}}/(1+T/N)$, where $\lambda_{\mathrm{LS}}$ is the population least-squares slope of $Y$ on $p(X)$, with an intercept. PPI++ targets the full variance within a scalar family \citep{angelopoulos2024ppiplusplus}. Remark~\ref{rem:variance_targets} in Appendix~\ref{app:full_variance_fit} gives the general known-representer formula. That appendix develops direct full-variance fitting, and Section~\ref{sec:stacking} optimizes the full variance by combining predictors.

\subsection{Stacking predictors}
\label{sec:stacking}

Stacking forms a weighted average of estimates from predictors fitted on common folds. We choose nonnegative weights summing to $1$ to minimize its estimated variance. These weights use the evaluation labels, so exact unbiasedness need not hold. Appendix~\ref{app:stacking} establishes valid inference under the stated convergence and nondegeneracy conditions.

\section{Application: the effects of an unconditional cash transfer}
\label{sec:application}

\paragraph{The experiment.} ORUS is the largest randomized trial of an unconditional cash transfer in the United States \citep{vivalt2024,bartik2024}. It enrolled 3,000 low-income adults aged 21--40 in Illinois and Texas and randomized 1,000 of them to receive \$1,000 per month for three years and 2,000 to receive \$50 per month. We study the fourteen outcomes on expenditure, employment and income, and financial health and saving whose full-sample effect is distinguishable from zero (Appendix~\ref{app:outcomes}).

\paragraph{The design.} We assess precision with scarce labels in 500 resampling replicates. Within each arm, we draw labeled and target records independently with replacement from the fixed study cohort. Their sizes are 10\% and 90\% of that arm's record count, rounded to integers. Only labeled draws supply outcomes. We estimate each arm's mean using representer $1$ and subtract the control estimate from the treated estimate (Appendix~\ref{app:procedure}). Methods evaluated on the same cohort use the same draws.

We report each estimator's standard deviation across replicates, denoted SE, and variance reduction $r:=1-(\mathrm{SE}/\mathrm{SE}_{\mathrm{floor}})^2$, where $\mathrm{SE}_{\mathrm{floor}}$ is the standard deviation of the labeled-only difference in means. Coverage is the fraction of intervals $\hth\pm1.96\,\widehat V^{1/2}$ containing the full-cohort difference in means, $\Delta_n$, the target of this conditional resampling experiment. The plug-in variance includes both labeled and target sampling variation.

\paragraph{The predictions.} The AI system is Claude Opus~4.6 at temperature zero. From baseline survey responses, it predicts each participant's outcome under the \$1,000 and \$50 payments (\rev{for related uses of LLM-generated responses combined with human data, see \citealp{zhang2025agentic,wang2026llm}}), giving $p(1,w)$ and $p(0,w)$, respectively. Outcomes and realized treatment assignments of prediction-cohort participants are withheld. We obtain two versions of $p$: a zero-shot prompt without worked examples and a few-shot prompt with five held-out examples for the specified payment. These examples are excluded from the few-shot analysis cohorts. Further prompts, based on liquidity constraints and labor supply, return AI-generated features (Appendix~\ref{app:prompts}).

\paragraph{The estimators.} All use the armwise version of \eqref{eq:estimator} or \eqref{eq:theta_cf} and differ in the predictor. \emph{Fixed PPI} sets $f=p$ with the zero-shot prediction without fitting any parameters. \emph{PPI++} estimates a separate scalar coefficient in each arm, setting $f(d,w)=\widehat\lambda_d p(d,w)$.

We consider four classes of predictors, each fitted with ten-fold cross-fitting. M1 calibrates the zero-shot prediction with arm intercepts, for example $f(d,w)=b_d+c\,p(d,w)$ or an isotonic map of $p$. M2 combines zero-shot predictions with baseline covariates, as in $f(d,w)=b_d+c\,p(d,w)+\beta^{\top}h(d,w)$, where $h$ collects the covariates. M3 uses few-shot predictions with covariates, and M4 incorporates AI-generated features. M1 contains linear, isotonic, and boosted-tree learners; M2--M4 also include shrunken slopes and ridge. Prediction loss fits the individual learners, and stacking chooses their weights in each replicate to minimize estimated variance. Appendix~\ref{app:full} compares the learner-combination rules and fitting criteria.

\begin{figure}[htbp]
\centering
\includegraphics[width=0.90\linewidth]{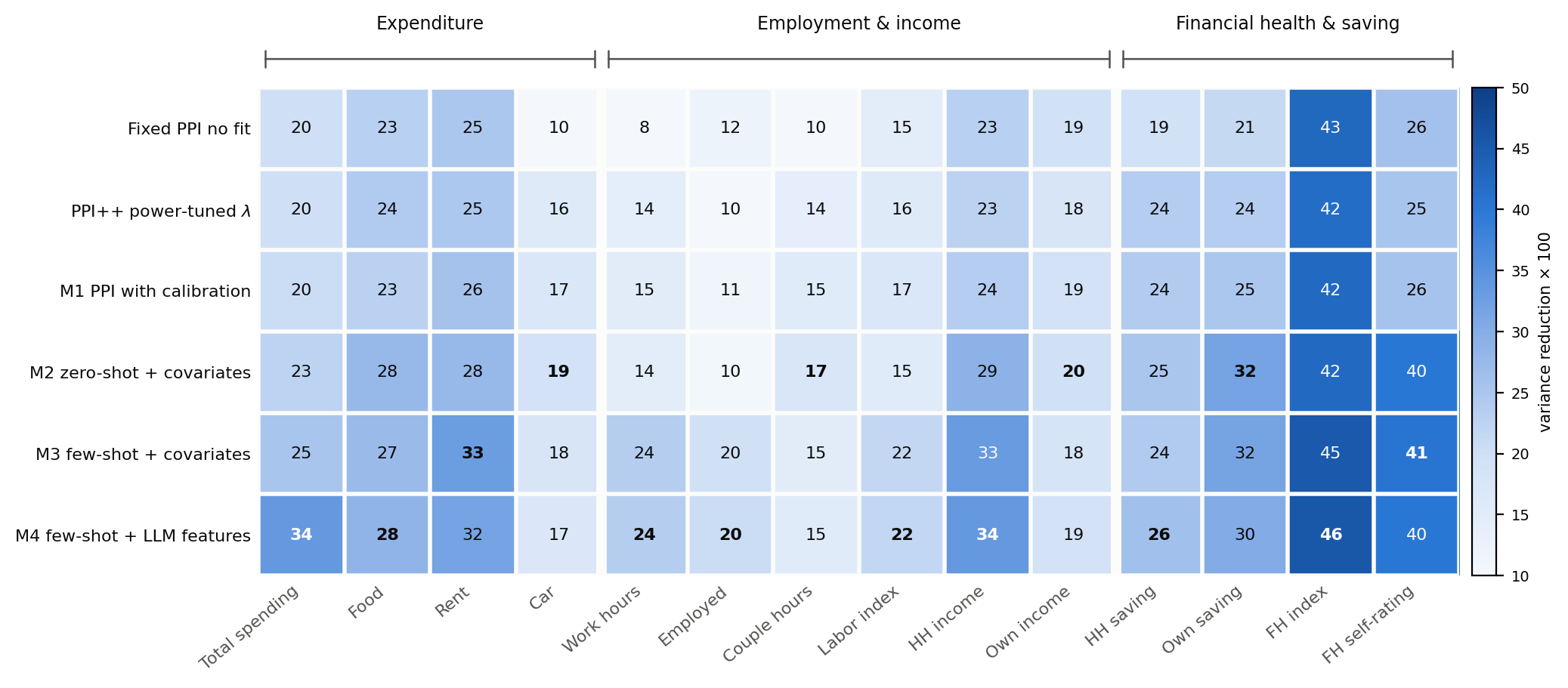}
\vspace{-7pt}
\caption{Variance reduction ($100r$), with 10\% labeled per arm. Each cell uses its cohort's labeled-only reference; M3--M4 exclude few-shot examples. M1--M4 are cross-fitted and stacked. Bold marks their best class before rounding.}
\label{fig:heatmap}
\vspace{-7pt}
\end{figure}

\paragraph{Precision and coverage.} Figure~\ref{fig:heatmap} shows precision gains for every method and outcome. Fixed PPI reduces variance by 8.5--42.7\%, and M4 by 15.3--45.6\%. M4 has the largest average gain, 27.6\% compared with 19.6\% for fixed PPI (Table~\ref{tab:class_summary}). The best input class varies across outcomes; the bold cells describe this variation, not a selection rule used within a replicate.

Mean coverage ranges from 94.7\% to 94.8\% for M1--M4, compared with 95.0\% for fixed PPI. Appendix~\ref{app:full} gives outcome-level standard errors and coverage, class summaries, and learner-combination comparisons.

\begin{table}[t!]
\centering
\vspace{-7pt}
\caption{Mean variance reduction and coverage (\%) across fourteen outcomes, with 10\% labels per arm. M1--M4 use cross-fitted stacking.}
\label{tab:class_summary}
\vspace{-7pt}
\small
\begin{tabular}{@{}lrr@{}}
\toprule
Method & Variance reduction (\%) & Coverage (\%) \\
\midrule
Fixed PPI, $f=p$ & 19.6 & 95.0 \\
PPI++ & 21.0 & 94.9 \\
M1: calibrated zero-shot $p$ & 21.7 & 94.8 \\
M2: zero-shot $p$ and covariates & 24.4 & 94.8 \\
M3: few-shot $p$ and covariates & 26.8 & 94.7 \\
M4: few-shot $p$, covariates and AI features & 27.6 & 94.7 \\
\bottomrule
\end{tabular}
\end{table}

\paragraph{Inputs and learner combination.} Average variance reduction increases from M1 to M4. These comparisons also change learner menus and sometimes cohorts, so they do not isolate individual input contributions. Across the 42 outcome--class cells in M2--M4, stacking improves mean variance reduction by 0.9 percentage points over selecting the learner with the smallest estimated variance and 4.3 points over equal weights (Appendix~\ref{app:full}, Table~\ref{tab:stack}). Ridge is the best individual learner on average and trails stacking by only 0.3 percentage points.

\paragraph{The fitting objective.} All three fitting criteria yield similar precision, with a slight average advantage for prediction loss (Appendix~\ref{app:full}, Table~\ref{tab:objective}). Standard errors retain both variance components under every fitting rule.

\section{Conclusion}
\label{sec:conclusion}

Known design information makes prediction errors exactly correctable and lets us fit AI-assisted predictors for precision. Under our conditions, cross-fitting permits misspecified limits, while direct variance fitting attains the best leading variance in the chosen class when labels are asymptotically scarce. The ORUS resampling experiment illustrates potential gains, but broader validation is needed: it uses one trial, one language model, and one seed, selects outcomes by full-cohort significance, and uses a model trained after the study's results became public.

\clearpage
\bibliography{references}

\clearpage
\appendix
\begingroup
\section*{Guide to the appendices}
\label{app:guide}

Appendix~\ref{app:related} expands the literature discussion in Section~\ref{sec:intro}. Appendices~\ref{app:framework}--\ref{app:proof_construction} follow Sections~\ref{sec:framework}--\ref{sec:construction}. Appendix~\ref{app:framework} proves the fixed-predictor results and verifies the representer and continuity assumptions for all five examples in Table~\ref{tab:examples}. Appendix~\ref{app:finetune} proves the cross-fitting results and gives conditions for inference without cross-fitting based on entropy or algorithmic stability. Appendix~\ref{app:proof_construction} follows the four predictor-construction subsections: prediction classes and the representer, prediction-loss fitting, variance optimization, and stacking.

Appendix~\ref{app:empirical} collects the empirical materials in the order needed to understand and reproduce the experiment. Sections~\ref{app:outcomes}--\ref{app:procedure} describe the outcomes, AI prompts and generated features, learners and their inputs, and simulation procedure. Section~\ref{app:full} supplies the detailed 10\% results and comparisons of learner-combination rules and fitting criteria. Readers seeking outcome-level standard errors and benchmark coverage can turn directly to that subsection.

\subsection*{Contents}
\newcommand{\appcontentsline}[3][0pt]{%
  \par\noindent\hspace*{#1}\hyperref[#2]{\ref*{#2}\quad #3}%
  \nobreak\dotfill\nobreak\hyperref[#2]{\pageref*{#2}}\par}
\noindent\hfill Page\par\smallskip
\appcontentsline{app:related}{\textbf{Related literature}}
\appcontentsline{app:framework}{\textbf{Framework and fixed-predictor inference}}
\appcontentsline[1em]{app:proof_framework}{Proofs of the results in Section~\ref*{sec:framework}}
\appcontentsline[1em]{app:verify}{Verification of assumptions in the examples}
\appcontentsline[1em]{app:derivative_regularity}{Additional regularity for average derivatives}
\appcontentsline{app:finetune}{\textbf{Inference with fitted predictors}}
\appcontentsline[1em]{app:proof_crossfitting}{Proofs of the cross-fitting results}
\appcontentsline[1em]{app:insample_validity}{Inference without cross-fitting}
\appcontentsline[1em]{app:insample_entropy}{Control through entropy}
\appcontentsline[1em]{app:insample_stability}{Control through algorithmic stability}
\appcontentsline{app:proof_construction}{\textbf{Building the predictor}}
\appcontentsline[1em]{app:prediction_classes}{Prediction classes and the representer}
\appcontentsline[1em]{app:prediction_loss}{Fitting by prediction loss}
\appcontentsline[1em]{app:variance_construction}{Fitting for estimator variance}
\appcontentsline[2em]{app:variance_fit}{Residual-variance optimization}
\appcontentsline[2em]{app:full_variance_fit}{Full-variance optimization at general sample-size ratios}
\appcontentsline[1em]{app:stacking}{Stacking and selection}
\appcontentsline{app:empirical}{\textbf{Empirical materials and additional results}}
\appcontentsline[1em]{app:outcomes}{Outcomes and their construction}
\appcontentsline[1em]{app:prompts}{Prompts and AI-generated features}
\appcontentsline[1em]{app:learners}{The learner menu and its inputs}
\appcontentsline[1em]{app:procedure}{The simulation procedure}
\appcontentsline[1em]{app:full}{Additional results}
\par\endgroup
\clearpage

\section{Related literature}
\label{app:related}

We build on the orthogonal corrections and cross-fitting methods of double/debiased machine learning \citep{chernozhukov2018dml}, including their formulation for linear functionals using Riesz representers \citep{chernozhukov2022riesz,chernozhukov2025covshift}. In formulations that estimate the representer, its estimation error interacts with the regression error and must be controlled. Here the representer is known, so this source of error is absent and the correction remains exact for a possibly misspecified predictor. Under independent sampling and integrability, cross-fitting preserves this property after fitting: conditional on the training observations, the evaluation observations retain their original distributions, and \eqref{eq:intro_identity} applies to the realized predictor. Under the stated moment and continuity conditions, we establish valid inference when the fitted predictors stabilize at a deterministic limit, which may be misspecified, without imposing a regression convergence rate.

The framework builds on the prediction-plus-correction principle of prediction-powered inference \citep[PPI;][]{angelopoulos2023science}. For a mean without distribution shift, replacing the expectations in \eqref{eq:intro_identity} by sample averages gives the PPI estimator. We develop this principle across the functionals in Table~\ref{tab:examples}, using their known representers to determine both the correction and the variance objective. PPI++ already optimizes a scalar adjustment to a fixed prediction for asymptotic variance \citep{angelopoulos2024ppiplusplus}; our direct fitting and stacking procedures develop variance optimization over possibly richer, misspecified prediction classes. Related approaches to adapting predictions include cross-fitting within PPI \citep{zrnic2024cross}, isotonic calibration for means \citep{vanderlaan2026calibeating}, and learning an imputed loss through recalibration \citep{ji2025surrogates}.

For treatment effects, the framework also builds on augmented inverse-probability weighting \citep{robins1994estimation} and regression adjustment in experiments \citep{freedman2008regadj,lin2013agnostic}, with connections to prognostic-score adjustment \citep{schuler2022increasing} and PPI for clinical trials \citep{poulet2025bmc}. These are particular applications of the general correction, whose design-weighting principle goes back to \citet{horvitz1952ht}.

\section{Framework and fixed-predictor inference}
\label{app:framework}

We give the proofs for Section~\ref{sec:framework} and verify the representer and continuity assumptions for the examples in Table~\ref{tab:examples}. The continuity bounds also support inference with fitted predictors in Section~\ref{sec:finetune}.

\subsection{Proofs of the results in Section~\ref{sec:framework}}
\label{app:proof_framework}

\begin{proofof}{Lemma~\ref{lem:identity}}
Since $f,\gamma_0\in\mathcal L$ and $\mathcal L$ is linear, the difference $\Delta:=f-\gamma_0$ belongs to $\mathcal L$. Assumption~\ref{ass:known_riesz} makes the target terms integrable. The square-integrability of $Y$, $\alpha_0(X)$, $f(X)$, and $\gamma_0(X)$ ensures, by Cauchy--Schwarz, that their weighted products $\alpha_0(X)Y$, $\alpha_0(X)f(X)$, and $\alpha_0(X)\gamma_0(X)$ are integrable. We may therefore condition the weighted residual on $X$:
\[
\E_P\,\alpha_0(X)\{Y-\gamma_0(X)\}
=\E_P\bigl[\alpha_0(X)\{\E_P[Y\mid X]-\gamma_0(X)\}\bigr]=0.
\]
By linearity and \eqref{eq:representer},
\[
\E_Q\,m(Z,f)-\theta_0
=\E_Q\,m(Z,\Delta)
=\E_P\,\alpha_0(X)\Delta(X).
\]
The weighted-residual expectation is the negative of the last term, which proves \eqref{eq:identity}.
\end{proofof}

\begin{proofof}{Lemma~\ref{lem:unbiased_var}}
The summands in \eqref{eq:estimator} are integrable by Assumption~\ref{ass:known_riesz} and the argument above. Taking expectations and applying Lemma~\ref{lem:identity} gives $\E\,\hth(f)=\theta_0$, using only the marginal laws $P$ and $Q$. When the summand variances are finite, independence within and between the samples gives $\Var\,\hth(f)=\sigma_Q^2(f)/N+\sigma_P^2(f)/T$.
\end{proofof}

\begin{proofof}{Theorem~\ref{thm:clt}}
Let $M_i:=m(Z_i,f)$ and $G_t:=\alpha_0(X_t)\{Y_t-f(X_t)\}$, with means $\mu_Q$ and $\mu_P$ and variances $\sigma_Q^2$ and $\sigma_P^2$; the argument $f$ is suppressed. By Lemma~\ref{lem:identity}, $\theta_0=\mu_Q+\mu_P$, so that
\[
\hth(f)-\theta_0=(\bar M_N-\mu_Q)+(\bar G_T-\mu_P),
\]
where $\bar M_N$ and $\bar G_T$ are the two sample means.

If either component variance is zero, its summands are constant almost surely, and its centered sample mean vanishes. The other component has positive variance because $\sigma^2(f)>0$, so the ordinary central limit theorem proves the first claim in this case. We therefore first establish the normal approximation when both component variances are positive.

Define the standardized sample means
\[
U_N:=\frac{\sqrt N(\bar M_N-\mu_Q)}{\sigma_Q},
\qquad
V_T:=\frac{\sqrt T(\bar G_T-\mu_P)}{\sigma_P}.
\]
The classical central limit theorem applies to each sample mean because its summands are independent and identically distributed with finite variance. Independence between the samples makes the joint characteristic function factorize, giving
\[
(U_N,V_T)\toD(\zeta_1,\zeta_2),
\]
where $\zeta_1$ and $\zeta_2$ are independent standard normal variables.

To combine the two sample means at the estimator's standard-error scale, define
\[
\lambda_Q:=\frac{\sigma_Q}{\sqrt N\,\sigma},
\qquad
\lambda_P:=\frac{\sigma_P}{\sqrt T\,\sigma},
\qquad
\sigma:=\sigma(f).
\]
These deterministic coefficients satisfy $\lambda_Q^2+\lambda_P^2=1$, and
\[
\frac{\hth(f)-\theta_0}{\sigma}=\lambda_QU_N+\lambda_PV_T.
\]

The coefficients need not converge because no restriction is imposed on $N/T$. We handle this by considering an arbitrary subsequence of sample sizes. Since $\lambda_Q\in[0,1]$, it has a further subsequence along which $\lambda_Q\to\bar\lambda_Q$, and then $\lambda_P\to\bar\lambda_P:=(1-\bar\lambda_Q^2)^{1/2}$. Along this further subsequence, joint convergence of the standardized means gives
\[
\lambda_QU_N+\lambda_PV_T
\toD\bar\lambda_Q\zeta_1+\bar\lambda_P\zeta_2
\sim\mathcal N(0,1).
\]
The limiting variance is $1$ because the normal variables are independent and the squared coefficients sum to $1$. Every subsequence therefore has a further subsequence with the same standard normal limit, which proves convergence of the full sequence.

It remains to justify replacing the variance by its sample estimate. The law of large numbers applied to the summands and their squares gives $\widehat\sigma_Q^2\toP\sigma_Q^2$ and $\widehat\sigma_P^2\toP\sigma_P^2$. Only second moments are needed for these applications. A component with zero population variance also has zero sample variance almost surely. Consequently, the relative error of the estimated total variance is bounded by the largest relative error among its positive components:
\[
\left|\frac{\widehat V(f)}{\sigma^2(f)}-1\right|
\le
\max_{\substack{j\in\{Q,P\}\\ \sigma_j^2>0}}
\left|\frac{\widehat\sigma_j^2}{\sigma_j^2}-1\right|
\toP0.
\]
This bound applies whether one or both component variances are positive. It also implies that $\widehat V(f)>0$ with probability tending to $1$. Slutsky's lemma now gives the studentized normal approximation.
\end{proofof}

\begin{proofof}{Proposition~\ref{prop:oracle}}
We first verify that the true regression is an admissible predictor for the variance minimization. Conditional Jensen's inequality gives
\[
\E_P[\alpha_0(X)^2\gamma_0(X)^2]
\le
\E_P\!\left[\alpha_0(X)^2\E_P[Y^2\mid X]\right]
=
\E_P[\alpha_0(X)^2Y^2]<\infty.
\]
Thus $\alpha_0(X)\{Y-\gamma_0(X)\}$ is square-integrable, and $\sigma_P^2(\gamma_0)<\infty$. Since $\gamma_0\in\mathcal L$, it belongs to the class over which the variance is minimized.

For admissible $f$, let $G_f:=\alpha_0(X)\{Y-f(X)\}$. Its conditional mean is $\E_P[G_f\mid X]=\alpha_0(X)\{\gamma_0(X)-f(X)\}$, and its conditional variance is $\Var_P(G_f\mid X)=\alpha_0(X)^2\Var_P(Y\mid X)$. The latter does not depend on $f$. By the law of total variance,
\[
\sigma_P^2(f)
=
\E_P\!\left[\alpha_0(X)^2\Var_P(Y\mid X)\right]
+
\Var_P\!\left(\alpha_0(X)\{\gamma_0(X)-f(X)\}\right).
\]
At $f=\gamma_0$, the conditional mean of $G_f$ is zero, so the first term equals $\sigma_P^2(\gamma_0)$. The second term is nonnegative and vanishes at $f=\gamma_0$, which proves both the variance decomposition and the minimizing property.
\end{proofof}

\subsection{Verification of assumptions in the examples}
\label{app:verify}

The following calculations verify Assumptions~\ref{ass:known_riesz} and~\ref{ass:msc} for the five examples in Table~\ref{tab:examples}. Assumption~\ref{ass:msc} is stated for a fixed class $\mathcal D$ and asks that the deviations $\hf_k-\fstar$ lie in it with probability tending to $1$. Membership can be enforced by construction, for instance by clipping predictions or by restricting the fine-tuning family.

\paragraph{Example 1 (mean without covariate shift).} Here $Z=X$, $Q=P_X$, and $m(z,f)=f(z)$. The representer identity holds with $\alpha_0=1$, and $\E_Qm(Z,\Delta)^2=\nP{\Delta}^2$. Thus Assumption~\ref{ass:known_riesz} holds, and Assumption~\ref{ass:msc} holds with $a=2$, $M=1$, and $\mathcal D=\mathcal L$. The membership condition is automatic because $\hf_k$ and $\fstar$ belong to the linear space $\mathcal L$.

\paragraph{Example 2 (treatment effect in an RCT).} Here $X=(D,W)$, $Z=W$, $m(w,f)=f(1,w)-f(0,w)$, and $D$ given $W=w$ is Bernoulli with known probability $\pi(w)$. For square-integrable $\Delta$,
$\E_P[\alpha_0(D,W)\Delta(D,W)\mid W]=\Delta(1,W)-\Delta(0,W)$, and taking expectations gives \eqref{eq:representer}. A known density ratio $\varpi(w)$ between the target and the labeled covariate distributions multiplies $\alpha_0$ by $\varpi(W)$.

For the following continuity bounds, we take $Q$ to be the distribution of $W$ under $P$ and $\alpha_0$ to be the RCT representer in Table~\ref{tab:examples}. There are two routes to Assumption~\ref{ass:msc}. Under uniform overlap, $\varepsilon\le\pi(W)\le1-\varepsilon$, we have
$m(W,\Delta)^2\le2\{\Delta(1,W)^2+\Delta(0,W)^2\}\le(2/\varepsilon)\,\E_P[\Delta(X)^2\mid W]$, so the assumption holds with $a=2$, $M=2/\varepsilon$ and $\mathcal D=\mathcal L$. Without uniform overlap, suppose only that $\E_P\,\alpha_0(X)^2<\infty$ and take $\mathcal D:=\{\Delta\in\mathcal L:|\Delta|\le B\}$. Since $|m(W,\Delta)|\le2B$,
\[
m(W,\Delta)^2=\bigl|\E_P[\alpha_0(X)\Delta(X)\mid W]\bigr|\cdot|m(W,\Delta)|\le2B\,\E_P\bigl[|\alpha_0(X)\Delta(X)|\bigm| W\bigr],
\]
and the Cauchy--Schwarz inequality gives
$\E_Q\,m(W,\Delta)^2\le2B\,(\E_P\,\alpha_0(X)^2)^{1/2}\,\nP{\Delta}$. The assumption holds with $a=1$.

\paragraph{Example 3 (average derivative).} Here $Z=X=(D,W)$, $Q=P_X$, and $m((d,w),f)=\partial_df(d,w)$. The known conditional density $\varrho(d\mid w)$ gives the representer $\alpha_0(d,w)=-\partial_d\log\varrho(d\mid w)$ by integration by parts. Appendix~\ref{app:derivative_regularity} supplies the domain, boundary, and moment conditions for this identity. For deviations satisfying the conditions of Lemma~\ref{lem:reverse_poincare}, the weak reverse Poincar\'e inequality and the uniform bound~\eqref{eq:derivative_uniform_k} give $\E_Qm(Z,\Delta)^2\le M\nP{\Delta}$ on a fixed class $\mathcal D$. This is the continuity bound in Assumption~\ref{ass:msc} with $a=1$, and the fitted-deviation membership condition remains required.

\paragraph{Example 4 (dose shift).} The target compares the outcome under a fixed shift $D\mapsto D+\delta$ with the outcome under the original dose. Here $X=Z=(D,W)$, $Q=P_X$, and $m((d,w),f)=f(d+\delta,w)-f(d,w)$. For almost every $w$, assume that the conditional distribution of $D+\delta$ given $W=w$ is absolutely continuous with respect to that of $D$ given $W=w$. This condition ensures that the shifted distribution places no mass on sets having zero probability under the observed distribution. Define
\[
r_\delta(d,w):=\frac{\varrho(d-\delta\mid w)}{\varrho(d\mid w)},
\qquad
\alpha_0(d,w):=r_\delta(d,w)-1,
\]
where $\varrho$ is the known conditional density or probability mass function, taken as zero outside its support. The ratio is defined $P_X$-almost surely, and we assume $\E_Pr_\delta(X)^2<\infty$.

A conditional change of variables, or reindexing of the sum for a discrete dose, gives $\E_Ph(D+\delta,W)=\E_P[r_\delta(X)h(X)]$ for every nonnegative measurable $h$. For $\Delta\in\mathcal L$, apply this identity to $h=|\Delta|$ and use Cauchy--Schwarz with the square-integrable functions $r_\delta$ and $\Delta$. The shifted evaluation is therefore integrable, and the identity extends to $\Delta$, giving
\[
\begin{aligned}
\E_Qm(Z,\Delta)
&=\E_P\{\Delta(D+\delta,W)-\Delta(D,W)\}\\
&=\E_P\{[r_\delta(X)-1]\Delta(X)\}.
\end{aligned}
\]
The function $\alpha_0=r_\delta-1$ is known and square-integrable, so Assumption~\ref{ass:known_riesz} holds. This argument requires no differentiability of the predictor and applies to both continuous and discrete doses.

To verify mean-square continuity, apply $(u-v)^2\le2u^2+2v^2$ and the same change of distribution to obtain
\[
\E_Qm(Z,\Delta)^2
\le 2\E_P\!\left[(r_\delta(X)+1)\Delta(X)^2\right].
\]
If $r_\delta\le\bar r_\delta<\infty$, this is at most $2(\bar r_\delta+1)\nP{\Delta}^2$. Assumption~\ref{ass:msc} therefore holds with $a=2$, $M=2(\bar r_\delta+1)$, and $\mathcal D=\mathcal L$. If the ratio is unbounded, take instead the fixed class $\mathcal D:=\{\Delta\in\mathcal L:|\Delta|\le B\}$, where the bound holds $P_X$-almost surely. Then
\[
\begin{aligned}
\E_Qm(Z,\Delta)^2
&\le 2B\,\E_P[(r_\delta(X)+1)|\Delta(X)|]\\
&\le 2B\,\nP{r_\delta+1}\nP{\Delta}.
\end{aligned}
\]
Thus the continuity bound in Assumption~\ref{ass:msc} holds with $a=1$ and $M=2B\nP{r_\delta+1}$.

\paragraph{Example 5 (mean under covariate shift).} Here $m(z,f)=f(z)$, and the target covariate distribution $Q$ may differ from $P_X$. Suppose $Q\ll P_X$, with known density ratio $r:=dQ/dP_X$ satisfying $\E_Pr(X)^2<\infty$. For every $\Delta\in\mathcal L$, Cauchy--Schwarz gives
\[
\E_Q|\Delta(Z)|
=\E_P[r(X)|\Delta(X)|]
\le\nP{r}\nP{\Delta}.
\]
The target evaluation is therefore integrable, and the change-of-measure identity $\E_Q\Delta(Z)=\E_P[r(X)\Delta(X)]$ verifies Assumption~\ref{ass:known_riesz} with $\alpha_0=r$.

To verify mean-square continuity, we use $\E_Qm(Z,\Delta)^2=\E_P[r(X)\Delta(X)^2]$. If $r\le\bar r<\infty$, Assumption~\ref{ass:msc} holds with $a=2$, $M=\bar r$, and $\mathcal D=\mathcal L$. If the ratio is unbounded, the class $\mathcal D:=\{\Delta\in\mathcal L:|\Delta|\le B\}$ gives
\[
\E_Qm(Z,\Delta)^2
\le B\,\E_P[r(X)|\Delta(X)|]
\le B\,\nP{r}\nP{\Delta}.
\]
This proves the continuity bound in Assumption~\ref{ass:msc} with $a=1$ and $M=B\nP{r}$.

A known ratio can arise from selective labeling. Suppose the target population has covariate distribution $Q$, and a unit with covariates $x$ is labeled with known probability $\tau(x)\ge\underline\tau>0$, independently of its outcome given $X$. The population labeling rate is $\bar\tau:=\E_Q\tau(X)$, and conditioning on being labeled gives $dP_X/dQ=\tau/\bar\tau$. If $\bar\tau$ is also known, then $r(x)=\bar\tau/\tau(x)$, which is bounded. For strata with known population shares $q_s$ and labeling probabilities $\tau_s$, the normalization is $\bar\tau=\sum_s q_s\tau_s$. Conditional independence of labeling and the outcome preserves the regression function across the labeled and target populations.

For dose shifts and means under covariate shift, the inference results also require finite summand variances and the stated conditions on the fitted predictors. A convenient sufficient condition for the variance requirements is that predictors are uniformly bounded by $B_0$ and $\E_P[Y^2\mid X]\le C_Y<\infty$. Then
\[
\E_P\!\left[\alpha_0(X)^2\{Y-f(X)\}^2\right]
\le 2(C_Y+B_0^2)\,\E_P\alpha_0(X)^2<\infty,
\]
and the target summand is bounded in both examples. For cross-fitted inference, weighted pseudo-consistency, membership of the fitted deviations in $\mathcal D$, and the nondegeneracy conditions remain as stated in Theorem~\ref{thm:cf_clt}.

\subsection{Additional regularity for average derivatives}
\label{app:derivative_regularity}

For an average derivative, integration by parts serves two purposes. It identifies the known representer, and a second application bounds the mean-square derivative of a prediction error by the size of that error. We give the conditions for each step, following the weak reverse Poincar\'e argument of \citet{chernozhukov_newey_singh_2023_biometrika}.

Let $Z=X=(D,W)$ and $Q=P_X$. For almost every $w$, suppose that $D\mid W=w$ has a known positive, continuously differentiable density $\varrho(d\mid w)$ on an interval $I=(\underline d,\bar d)$, whose endpoints may be infinite. Define
\[
\alpha_0(d,w):=-\partial_d\log\varrho(d\mid w)
\]
and assume $\nP{\alpha_0}<\infty$. All functions and derivatives below are assumed jointly measurable. Take $\mathcal L$ to be a linear space of square-integrable functions $f$ such that $f(\cdot,w)$ is locally absolutely continuous on $I$, $\E_P|\partial_df(D,W)|<\infty$, and
\begin{equation}
\label{eq:derivative_boundary}
\lim_{d\downarrow\underline d}f(d,w)\varrho(d\mid w)
=\lim_{d\uparrow\bar d}f(d,w)\varrho(d\mid w)=0
\end{equation}
for almost every $w$. As in Section~\ref{sec:framework}, assume $\gamma_0\in\mathcal L$. The boundary condition removes the endpoint terms when we transfer a derivative from $f$ to the density.

Indeed, for $f\in\mathcal L$, Cauchy--Schwarz gives $\E_P|\alpha_0(X)f(X)|<\infty$. Since $\partial_d\varrho=-\alpha_0\varrho$, this bound and $\E_P|\partial_df(D,W)|<\infty$ ensure that both $\partial_df\varrho$ and $f\partial_d\varrho$ are absolutely integrable over $I$ for almost every $w$. For each such $w$, we integrate by parts on compact subintervals of $I$ and let their endpoints approach those of $I$. Absolute convergence and the boundary condition~\eqref{eq:derivative_boundary} give
\begin{align*}
\int_I \partial_df(d,w)\varrho(d\mid w)\,\mathrm d d
&=-\int_I f(d,w)\partial_d\varrho(d\mid w)\,\mathrm d d\\
&=\int_I \alpha_0(d,w)f(d,w)\varrho(d\mid w)\,\mathrm d d.
\end{align*}
Both sides are absolutely integrable over $w$, so averaging gives
\begin{equation}
\label{eq:derivative_riesz}
\E_P\,\partial_df(D,W)=\E_P\,\alpha_0(X)f(X),
\qquad f\in\mathcal L.
\end{equation}
Thus Assumption~\ref{ass:known_riesz} holds for the derivative map in Table~\ref{tab:examples}, and Lemma~\ref{lem:identity} applies. The assumption $\E_PY^2<\infty$ also ensures that $\alpha_0(X)Y$ is integrable.

To verify mean-square continuity, we need control of the squared derivative of a deviation. The following result supplies that control through a second integration by parts, with its boundary term stated explicitly.

\begin{lemma}[Weak reverse Poincar\'e inequality]
\label{lem:reverse_poincare}
Under the density conditions above, let $f\in\mathcal L$ be twice continuously differentiable in $d$ for almost every $w$. Suppose
\begin{equation}
\label{eq:derivative_energy_boundary}
\lim_{d\downarrow\underline d} f(d,w)\partial_df(d,w)\varrho(d\mid w)
=\lim_{d\uparrow\bar d} f(d,w)\partial_df(d,w)\varrho(d\mid w)=0
\end{equation}
for almost every $w$, and define
\[
k_f(d,w):=\alpha_0(d,w)\partial_df(d,w)-\partial_d^2f(d,w).
\]
If $\nP{k_f}<\infty$, then
\begin{equation}
\label{eq:reverse_poincare}
\E_P\{\partial_df(D,W)\}^2
=\E_P[f(X)k_f(X)]
\le\nP{f}\,\nP{k_f}.
\end{equation}
\end{lemma}

\begin{proof}
Using $\partial_d\varrho=-\alpha_0\varrho$, the product rule gives
\[
\begin{aligned}
\partial_d\{\varrho(d\mid w)\partial_df(d,w)\}
&=\{\partial_d^2f(d,w)-\alpha_0(d,w)\partial_df(d,w)\}\varrho(d\mid w)\\
&=-k_f(d,w)\varrho(d\mid w).
\end{aligned}
\]
Fix $w$ for which the hypotheses hold and $\int_I|f k_f|\varrho\,\mathrm d d<\infty$; this includes almost every $w$ because $\E_P|f(X)k_f(X)|\le\nP{f}\nP{k_f}<\infty$. For $\underline d<u<v<\bar d$, integration by parts gives
\[
\int_u^v (\partial_df)^2\varrho\,\mathrm d d
=[f\partial_df\varrho]_u^v+\int_u^v f k_f\varrho\,\mathrm d d,
\]
where the arguments $(d,w)$ are suppressed. Let $u\downarrow\underline d$ and $v\uparrow\bar d$. The boundary term vanishes by \eqref{eq:derivative_energy_boundary}, the last integral converges absolutely, and the nonnegative integral on the left converges monotonically. Hence
\[
\int_I(\partial_df)^2\varrho\,\mathrm d d
=\int_I f k_f\varrho\,\mathrm d d.
\]
Tonelli's theorem on the left and absolute integrability on the right allow us to average over $W$. Cauchy--Schwarz then gives \eqref{eq:reverse_poincare}, including finiteness of the derivative's second moment.
\end{proof}

Apply the lemma to each $\Delta$ in a fixed class $\mathcal D\subseteq\mathcal L$ satisfying its smoothness and boundary conditions. If
\begin{equation}
\label{eq:derivative_uniform_k}
\sup_{\Delta\in\mathcal D}\nP{k_\Delta}\le M<\infty,
\end{equation}
then $\E_Qm(Z,\Delta)^2\le M\nP{\Delta}$. This is the bound in Assumption~\ref{ass:msc} with $a=1$. The class $\mathcal D$ need not be linear, and the bound is uniform over that class. For fitted deviations, one also requires $\Pp(\hf_k-\fstar\in\mathcal D\text{ for all }k)\to1$, as in that assumption.

A direct way to ensure \eqref{eq:derivative_uniform_k} is to impose fixed bounds
\[
\sup_{\Delta\in\mathcal D}\lVert\partial_d\Delta\rVert_\infty\le B_1,
\qquad
\sup_{\Delta\in\mathcal D}\nP{\partial_d^2\Delta}\le B_2,
\]
where $\lVert\cdot\rVert_\infty$ is the essential supremum under $P_X$. For every $\Delta\in\mathcal D$, the triangle inequality gives
\[
\nP{k_\Delta}
\le\nP{\alpha_0\partial_d\Delta}+\nP{\partial_d^2\Delta}
\le B_1\nP{\alpha_0}+B_2,
\]
so we may take $M=B_1\nP{\alpha_0}+B_2$. In particular, the representer may be unbounded. For a concrete bounded-dose design, take $I=(0,1)$ and $\varrho(d\mid w)=30d^2(1-d)^2$. Then
\[
\alpha_0(d,w)=-\frac2d+\frac2{1-d},
\qquad
\E_P\alpha_0(X)^2=120\int_0^1(2d-1)^2\,\mathrm d d=40.
\]
Uniformly bounded functions and first two dose derivatives satisfy both boundary conditions and the displayed bounds. Linear combinations of finitely many such smooth functions, with bounded coefficients, provide one possible tuning class; differences of its elements satisfy the same conditions with enlarged bounds.

For the variance and normal approximation results, the labeled residual still needs a second moment, $\E_P[\alpha_0(X)^2\{Y-f(X)\}^2]<\infty$. A bounded predictor and a uniformly bounded conditional second moment $\E_P[Y^2\mid X]$ suffice because $\nP{\alpha_0}<\infty$. The lemma supplies the target second moment whenever its conditions hold for $f$. For cross-fitted inference, the uniform bound above verifies mean-square continuity, while the weighted pseudo-consistency and moment conditions in Section~\ref{sec:finetune} remain in force.

\section{Inference with fitted predictors}
\label{app:finetune}

We first prove the cross-fitting results of Section~\ref{sec:finetune}. We then give conditions under which all labeled observations can be reused for fitting and evaluation, using entropy control or algorithmic stability.

\subsection{Proofs of the cross-fitting results}
\label{app:proof_crossfitting}

Throughout these proofs, let $\mathcal H_k$ denote the information used to train $\hf_k$: the labeled observations outside $I_k$ and any permitted independent external information. Conditional on $\mathcal H_k$, the fit is fixed, while the held-out observations and the target sample retain their original independent product laws. We use this conditioning separately for each fold; the fold estimates themselves need not be independent.

\begin{proofof}{Proposition~\ref{prop:cf_unbiased}}
Fix a fold $k$. For almost every realization of its training information, $\hf_k\in\mathcal L$, and Lemma~\ref{lem:identity} shows that the two summand means in $\hth_k$ add to $\theta_0$. Independence of the held-out and target samples therefore gives
\[
\E\!\left[(\hth_k-\theta_0)^2\mid\mathcal H_k\right]
=\frac{\sigma_Q^2(\hf_k)}{N}+\frac{\sigma_P^2(\hf_k)}{|I_k|}.
\]
Tonelli's theorem and the expected variance conditions imply that this squared error has finite expectation. Thus $\hth_k$ is integrable, and we may apply the tower property to
\[
\E[\hth_k\mid\mathcal H_k]
=\E_Q\,m(Z,\hf_k)+\E_P\!\left[\alpha_0(X)\{Y-\hf_k(X)\}\right]
=\theta_0.
\]
Here the expectations on the right integrate over a new observation with the fitted function held fixed. Taking expectations and averaging over folds with the deterministic weights $\omega_k$ proves the claim.
\end{proofof}

\begin{proofof}{Lemma~\ref{lem:quasi_oracle}}
Write $\Delta_k=\hf_k-\fstar$. The representer identity gives
\[
\mu_k:=\E_Q\,m(Z,\Delta_k)=\E_P[\alpha_0(X)\Delta_k(X)],
\]
where both expectations hold $\Delta_k$ fixed. Consequently, the population prediction errors cancel. By linearity of $m$,
\[
\hth_{\CF}-\hth(\fstar)=\sum_{k=1}^{K}\omega_k(A_k-C_k),
\]
where
\[
A_k:=\frac1N\sum_{i=1}^{N}\{m(Z_i,\Delta_k)-\mu_k\},
\qquad
C_k:=\frac1{|I_k|}\sum_{t\in I_k}\{\alpha_0(X_t)\Delta_k(X_t)-\mu_k\}.
\]
Conditional on $\mathcal H_k$, these are independent centered averages. Their conditional second moments satisfy
\[
\E[A_k^2\mid\mathcal H_k]\le\frac1N\,\E_Q\,m(Z,\Delta_k)^2,
\qquad
\E[C_k^2\mid\mathcal H_k]\le\frac{\nPa{\Delta_k}^2}{\underline\omega T}.
\]
To apply mean-square continuity, use the fold-specific event
\[
\mathcal E_k:=\{\Delta_k\in\mathcal D\}.
\]
This event is measurable with respect to $\mathcal H_k$, and Assumption~\ref{ass:msc} gives $\Pp(\bigcap_k\mathcal E_k)\to1$. On $\mathcal E_k$, the preceding bounds imply
\[
\E[(A_k-C_k)^2\mid\mathcal H_k]
\le\frac{M}{N}\nP{\Delta_k}^{a}
+\frac{1}{\underline\omega T}\nPa{\Delta_k}^{2}
\le C_0r_k^2,
\]
where
\[
r_k:=N^{-1/2}\nP{\Delta_k}^{a/2}+T^{-1/2}\nPa{\Delta_k},
\qquad C_0:=\max\{M,\underline\omega^{-1}\}.
\]
Each $r_k$ is determined by the training information. Conditional Chebyshev's inequality therefore yields
\[
\Pp\!\left(\mathcal E_k\cap\{|A_k-C_k|>ur_k\}\right)
\le\frac{C_0}{u^2},\qquad u>0.
\]
If $r_k=0$, the conditional second-moment bound makes $A_k-C_k=0$ on $\mathcal E_k$, so the same inequality applies.

Let $R$ be the scale appearing in the lemma:
\[
R:=N^{-1/2}\max_{k\le K}\nP{\Delta_k}^{a/2}
+T^{-1/2}\max_{k\le K}\nPa{\Delta_k}.
\]
Since $r_k\le R$ and the weights sum to one, a union bound gives
\[
\Pp\!\left(|\hth_{\CF}-\hth(\fstar)|>uR\right)
\le\Pp\!\left(\bigcup_{k=1}^{K}\mathcal E_k^c\right)+\frac{KC_0}{u^2}.
\]
The first term tends to zero, and $K$ is fixed. This proves the asserted stochastic order without requiring independence across folds.
\end{proofof}

\begin{proofof}{Theorem~\ref{thm:cf_clt}}
Write
\[
\hth_{\CF}-\theta_0
=\{\hth(\fstar)-\theta_0\}+\{\hth_{\CF}-\hth(\fstar)\}.
\]
Assumption~\ref{ass:pseudo} and the fixed number of folds imply
\[
\max_k\nPa{\Delta_k}=\op(1),
\qquad \max_k\nP{\Delta_k}^{a/2}=\op(1).
\]
Lemma~\ref{lem:quasi_oracle} thus makes the second term $\op(N^{-1/2}+T^{-1/2})$. Because both variance components at $\fstar$ are positive,
\[
\sigma(\fstar)
=\left\{\frac{\sigma_Q^2(\fstar)}{N}+\frac{\sigma_P^2(\fstar)}{T}\right\}^{1/2}
\ge c\,(N^{-1/2}+T^{-1/2})
\]
for some constant $c>0$. The fitting remainder is therefore $\op(\sigma(\fstar))$. Applying Theorem~\ref{thm:clt} to the deterministic predictor $\fstar$, followed by Slutsky's lemma, proves asymptotic normality.

For variance consistency, we first explain how to pass from conditional moment bounds to convergence in probability. Suppose $U\ge0$, and
\[
\E[U\mid\mathcal H]\le b\quad\text{on }\mathcal E,
\]
where $b\ge0$ and $\mathcal E$ are $\mathcal H$-measurable. Conditional Markov's inequality gives, for every $\varepsilon,\eta>0$,
\[
\Pp(U>\varepsilon)
\le\Pp(\mathcal E^c)+\Pp(b>\eta)+\frac{\eta}{\varepsilon}.
\]
Hence $U=\op(1)$ whenever $b=\op(1)$ and $\Pp(\mathcal E)\to1$. This implication does not require $\E b\to0$.

Apply this observation to the target-sample prediction errors. For each fold,
\[
U_{Q,k}:=\frac1N\sum_{i=1}^{N}m(Z_i,\Delta_k)^2
\]
has conditional expectation at most $M\nP{\Delta_k}^{a}=\op(1)$ on $\mathcal E_k$. Thus $U_{Q,k}=\op(1)$. Define
\[
a_i:=m(Z_i,\fstar),
\qquad d_i:=\sum_{k=1}^{K}\omega_k\,m(Z_i,\Delta_k).
\]
The target contributions entering the variance estimator are $a_i+d_i$. Jensen's inequality gives
\[
\frac1N\sum_{i=1}^{N}d_i^2\le\sum_{k=1}^{K}\omega_k U_{Q,k}=\op(1).
\]

For any two arrays $u,v$, the triangle inequality for the sample standard deviation yields
\[
\left|\widehat\Var_n(u+v)^{1/2}-\widehat\Var_n(u)^{1/2}\right|
\le\widehat\Var_n(v)^{1/2}.
\]
This is a deterministic inequality and requires no independence between the arrays. Since the empirical second moment of $d$ vanishes, its sample variance does too. The law of large numbers applied to $a_i$ and $a_i^2$ therefore gives
\[
\widehat\Var_N(a+d)\toP\sigma_Q^2(\fstar).
\]

For the labeled sample, write the fitted residual contributions as $g_t-e_t$, where
\[
g_t:=\alpha_0(X_t)\{Y_t-\fstar(X_t)\},
\qquad e_t:=\alpha_0(X_t)\Delta_{k(t)}(X_t).
\]
Within fold $k$,
\[
\E\!\left[\frac1{|I_k|}\sum_{t\in I_k}\alpha_0(X_t)^2\Delta_k(X_t)^2
\,\middle|\,\mathcal H_k\right]
\le\nPa{\Delta_k}^2=\op(1).
\]
Averaging the foldwise bounds gives $T^{-1}\sum_t e_t^2=\op(1)$ by the conditional Markov argument. Applying the same sample-standard-deviation inequality, together with the law of large numbers for the oracle residuals $g_t$, yields
\[
\widehat\Var_T(g-e)\toP\sigma_P^2(\fstar).
\]

Finally, positivity of both limiting components implies
\[
\left|\frac{\widehat V_{\CF}}{\sigma^2(\fstar)}-1\right|
\le\max\left\{
\left|\frac{\widehat\Var_N(a+d)}{\sigma_Q^2(\fstar)}-1\right|,
\left|\frac{\widehat\Var_T(g-e)}{\sigma_P^2(\fstar)}-1\right|
\right\}=\op(1).
\]
This proves relative variance consistency for arbitrary relative growth of $N$ and $T$.
\end{proofof}

\begingroup
\widowpenalty=10000
\subsection{Inference without cross-fitting}
\label{app:insample_validity}
\label{app:insample}

Cross-fitting separates each fit from the observations used to evaluate its correction. We now give conditions for valid inference when all labels are reused. Throughout Appendices~\ref{app:insample_validity}--\ref{app:insample_stability}, the labeled and target samples are independent, as in Section~\ref{sec:framework}, and the fit uses only the labeled sample. We first isolate the dependence that must be controlled. Let $\mathbb P_T$ and $\mathbb Q_N$ denote the labeled and target empirical measures, and write $Ph:=\E_Ph(X)$ and $Qg:=\E_Qg(Z)$. Let $\hf\in\mathcal L$ be a measurable fit based on all labeled observations, let $\fstar\in\mathcal L$ be deterministic, and define
\[
\Delta:=\hf-\fstar,
\qquad h:=\alpha_0\Delta,
\qquad \mathbb G_T:=\sqrt T(\mathbb P_T-P).
\]
For a realized random function, $P$ and $Q$ integrate only its argument, holding the function fixed. Subtracting the two estimators first gives
$\hth(\hf)-\hth(\fstar)=\mathbb Q_Nm(\cdot,\Delta)-\mathbb P_Th$.
The representer identity applies to each realized $\Delta\in\mathcal L$, so $Qm(\cdot,\Delta)=Ph$. Subtracting this common population mean from both empirical averages gives the exact decomposition
\begin{equation}
\label{eq:tradeoff}
\hth(\hf)-\hth(\fstar)
=(\mathbb Q_N-Q)m(\cdot,\Delta)
-(\mathbb P_T-P)h.
\end{equation}
The population terms cancel even if $\fstar$ differs from the regression function. Independence of the target sample controls the first empirical term. The second term requires control of the fit's dependence on its training observations.

\begin{proposition}[Inference without cross-fitting]
\label{prop:no_crossfit}
Suppose Assumption~\ref{ass:known_riesz} holds, the two samples are independent, and $\hf\in\mathcal L$ is measurable and fitted only on the labeled sample. Let $\fstar\in\mathcal L$ be deterministic, with $0<\sigma_Q^2(\fstar),\sigma_P^2(\fstar)<\infty$. Suppose, as $N,T\to\infty$, that
\begin{equation}
\label{eq:no_crossfit_conditions}
Q\{m(\cdot,\Delta)^2\}=\op(1),
\qquad \mathbb G_T h=\op(1),
\qquad \mathbb P_T h^2=\op(1).
\end{equation}
Then the estimator and plug-in variance from Section~\ref{sec:framework}, evaluated at $\hf$, satisfy
\[
\frac{\hth(\hf)-\theta_0}{\sigma(\fstar)}\toD\mathcal N(0,1),
\qquad
\frac{\widehat V(\hf)}{\sigma^2(\fstar)}\toP1,
\qquad
\frac{\hth(\hf)-\theta_0}{\widehat V(\hf)^{1/2}}\toD\mathcal N(0,1).
\]
\end{proposition}

The first condition follows from pseudo-consistency and mean-square continuity, as in Assumptions~\ref{ass:pseudo}--\ref{ass:msc}, with a single fit. The second is the stochastic equicontinuity condition needed for the point estimate. The third controls the empirical second moment of the change in weighted residuals, as needed for variance estimation. Population convergence alone does not justify this step when the fit is evaluated on its training data. These conditions give asymptotic validity, whereas cross-fitting also preserves exact unbiasedness under the integrability conditions of Proposition~\ref{prop:cf_unbiased}.

\begin{proofof}{Proposition~\ref{prop:no_crossfit}}
We first show that the fitting remainder in \eqref{eq:tradeoff} is negligible on the scale of the fixed-predictor estimator. Define
$A_N:=(\mathbb Q_N-Q)m(\cdot,\Delta)$ and $\rho_T:=Q\{m(\cdot,\Delta)^2\}$.
Conditional on the labeled sample, and on independent algorithmic randomization if used, $\Delta$ is fixed and the target observations retain their independent $Q$ laws. Thus $\E[A_N\mid\hf]=0$ and $\E[NA_N^2\mid\hf]\le\rho_T$.
To use $\rho_T=\op(1)$ without assuming that $\E\rho_T\to0$, fix $\varepsilon,\eta>0$. Conditional Chebyshev's inequality gives
\[
\Pp(|\sqrt N A_N|>\varepsilon)
\le\Pp(\rho_T>\eta)+\frac{\eta}{\varepsilon^2}.
\]
First let $N,T\to\infty$ and then let $\eta\downarrow0$. We obtain $A_N=\op(N^{-1/2})$. The second condition in \eqref{eq:no_crossfit_conditions} gives $(\mathbb P_T-P)h=\op(T^{-1/2})$. Combining the two bounds yields
\[
\hth(\hf)-\hth(\fstar)=\op(N^{-1/2}+T^{-1/2}).
\]
Since both component variances at $\fstar$ are positive, $\sigma(\fstar)$ is bounded below by a positive constant times $N^{-1/2}+T^{-1/2}$. The fitting remainder is therefore $\op(\sigma(\fstar))$, and Theorem~\ref{thm:clt} and Slutsky's lemma give the first conclusion.

We next compare the two estimated component variances with their fixed-predictor counterparts. Conditional on the fit, the nonnegative variable $\mathbb Q_N\{m(\cdot,\Delta)^2\}$ has expectation $\rho_T$. Conditional Markov's inequality, with the same truncation at $\rho_T\le\eta$, gives
$\mathbb Q_N\{m(\cdot,\Delta)^2\}=\op(1)$.
For any two arrays $(a_i)_{i\le n}$ and $(b_i)_{i\le n}$, the triangle inequality for their centered Euclidean norms implies
\[
\left|\widehat\Var_n(a+b)^{1/2}-\widehat\Var_n(a)^{1/2}\right|
\le\widehat\Var_n(b)^{1/2}
\le\left(\frac1{n-1}\sum_{i=1}^n b_i^2\right)^{1/2}.
\]
Because this inequality is deterministic, it applies even though the fitted residuals depend on all the labeled observations. Apply it first to the target summands, with $b_i=m(Z_i,\Delta)$, and then to the labeled residual summands, with $b_t=-h(X_t)$. The target bound just established and the third condition in \eqref{eq:no_crossfit_conditions} make both changes in sample standard deviation $\op(1)$. The law of large numbers applies to the fixed-predictor summands because their second moments are finite. Hence
\[
\widehat\sigma_Q^2(\hf)\toP\sigma_Q^2(\fstar),
\qquad
\widehat\sigma_P^2(\hf)\toP\sigma_P^2(\fstar).
\]
The ratio $\widehat V(\hf)/\sigma^2(\fstar)$ is a convex combination of the two component variance ratios, with weights proportional to $\sigma_Q^2(\fstar)/N$ and $\sigma_P^2(\fstar)/T$. Both ratios tend to $1$, so their convex combination does too, for every relative rate of $N$ and $T$. Slutsky's lemma now gives the studentized conclusion.
\end{proofof}

\subsection{Control through entropy}
\label{app:insample_entropy}

One way to control the training dependence is to bound the empirical process over a deterministic class that contains the fitted error with high probability. This is the approach to inference without sample splitting developed, for growing nuisance classes, by \citet{belloni2018uniform}. The following sufficient condition makes the trade-off between the size of the error and the complexity of its class explicit. It restricts the weighted error $h=\alpha_0(\hf-\fstar)$, which is the function evaluated in the labeled correction.

\begin{corollary}[An entropy condition]
\label{cor:no_crossfit_entropy}
Retain the setup, positive component variances, and target condition of Proposition~\ref{prop:no_crossfit}. Suppose $\Pp(h\in\mathcal H_T)\to1$, where $\mathcal H_T$ is a deterministic pointwise measurable class of functions satisfying $|g|\le B$ and $Pg^2\le r_T^2$ for all $g\in\mathcal H_T$, with a fixed $B>0$ and $0<r_T<B/2$. Suppose $v_T\ge1$, $A_T\ge e$, and
\[
\sup_{\nu}\log\mathcal N(\varepsilon B,\mathcal H_T,L^2(\nu))
\le v_T\log(A_T/\varepsilon),\qquad 0<\varepsilon\le1,
\]
where $\mathcal N$ is the covering number and the supremum ranges over finitely supported probability measures $\nu$ on $\mathcal X$. Define $L_T:=\log(A_TB/r_T)$. If
\begin{equation}
\label{eq:no_crossfit_entropy_rates}
r_T\sqrt{v_TL_T}+\frac{Bv_TL_T}{\sqrt T}\longrightarrow0,
\end{equation}
then all conclusions of Proposition~\ref{prop:no_crossfit} hold.
\end{corollary}

\begin{proofof}{Corollary~\ref{cor:no_crossfit_entropy}}
We must control both the centered empirical mean of $h$ and its empirical second moment. We start with the mean. For the class $\mathcal H_T$ with envelope $B$, define the uniform entropy integral
\[
J_T(\delta):=\sup_\nu\int_0^\delta
\sqrt{1+\log\mathcal N(\varepsilon B,\mathcal H_T,L^2(\nu))}\,d\varepsilon,
\qquad 0<\delta<1.
\]
The assumed covering bound gives
$J_T(\delta)\le C\delta\sqrt{v_T\log(A_T/\delta)}$ for a universal constant $C$.
Indeed, substituting $\varepsilon=\delta u$ separates $\log(A_T/\varepsilon)$ into $\log(A_T/\delta)+\log(1/u)$, and $\int_0^1\sqrt{\log(1/u)}\,du<\infty$.

Choose $\delta_T:=2r_T/B<1$ so that the normalized class $\mathcal H_T/B$, with envelope $1$, satisfies $P\{(g/B)^2\}\le\delta_T^2/4<\delta_T^2$, the strict moment bound required by the maximal inequality. Pointwise measurability supplies a countable subclass approximating every member pointwise, and the same approximation holds after squaring. Bounded convergence then makes the empirical-process suprema measurable. Thus the local maximal inequality of \citet[Theorem~2.1]{vanderVaartWellner2011local}, rescaled by $B$, yields
\begin{align*}
\E\sup_{g\in\mathcal H_T}|\mathbb G_Tg|
&\le CB\left\{J_T(\delta_T)
+\frac{J_T(\delta_T)^2}{\delta_T^2\sqrt T}\right\}\\
&\le C\left(r_T\sqrt{v_TL_T}+\frac{Bv_TL_T}{\sqrt T}\right)=o(1).
\end{align*}
The constant $C$ may increase between inequalities, but does not depend on $T$, $r_T$, $v_T$, $A_T$, or $B$. On the event $h\in\mathcal H_T$, $|\mathbb G_Th|$ is bounded by this supremum. Markov's inequality and $\Pp(h\notin\mathcal H_T)\to0$ therefore give $\mathbb G_Th=\op(1)$.

For the second moment, we apply the same argument to $\mathcal H_T^2:=\{g^2:g\in\mathcal H_T\}$. The inequality $|g^2-\widetilde g^2|\le2B|g-\widetilde g|$ implies
\[
\mathcal N(\varepsilon B^2,\mathcal H_T^2,L^2(\nu))
\le\mathcal N(\varepsilon B/2,\mathcal H_T,L^2(\nu)).
\]
Thus the squared class satisfies the covering bound with envelope $B^2$ and $A_T$ replaced by $2A_T$. Its second moments satisfy $Pg^4\le B^2r_T^2$, so its normalized radius is again $\delta_T$. The class $\{g^4:g\in\mathcal H_T\}$ is pointwise measurable as well, so the squared class also meets both measurability requirements of the theorem. The maximal inequality, followed by division by $\sqrt T$, gives
\[
\E\sup_{g\in\mathcal H_T}|(\mathbb P_T-P)g^2|
\le C\left(\frac{Br_T\sqrt{v_TL_T}}{\sqrt T}
+\frac{B^2v_TL_T}{T}\right)=o(1).
\]
Condition~\eqref{eq:no_crossfit_entropy_rates} also implies $r_T\to0$, since $v_TL_T\ge1$. On $h\in\mathcal H_T$, we now have
\[
\mathbb P_Th^2=Ph^2+(\mathbb P_T-P)h^2
\le r_T^2+\sup_{g\in\mathcal H_T}|(\mathbb P_T-P)g^2|.
\]
The right side tends to zero in probability, and the event has probability tending to $1$. This proves $\mathbb P_Th^2=\op(1)$, the remaining labeled-sample condition of Proposition~\ref{prop:no_crossfit}.
\end{proofof}

The complexity $v_T$ may grow with $T$, provided that \eqref{eq:no_crossfit_entropy_rates} holds. The condition therefore allows more than a fixed Donsker class. Boundedness is a sufficient condition for this particular corollary; it is not required by Proposition~\ref{prop:no_crossfit}.

\subsection{Control through algorithmic stability}
\label{app:insample_stability}

Alternatively, we can control how much the fitted weighted prediction changes when a training observation is replaced. This is the stability approach of \citet{chen2022stable}. We give a mean-square version that also controls the plug-in variance, with a proof specialized to the known-representer identity.

Write $O_t:=(Y_t,X_t)$ and $\mathcal S:=(O_1,\ldots,O_T)$, and let $O_1',\ldots,O_T'$ be independent copies, independent of $\mathcal S$. Let $\hf^{(t)}$ be the output of the same fitting rule when $O_t$ is replaced by $O_t'$. All fits use the same independent algorithmic randomization, if any. Define
\[
h_{\mathcal S}:=\alpha_0(\hf-\fstar),
\qquad h^{(t)}:=\alpha_0(\hf^{(t)}-\fstar).
\]
Let $X^\circ\sim P_X$ be independent of all training observations, their copies, and algorithmic randomization. To distinguish risk at a new observation from sensitivity at a training observation, define
\begin{align*}
r_T^2&:=\E\,P h_{\mathcal S}^2,\\
s_T&:=\max_{t\le T}\left(\E\,|h_{\mathcal S}(X_t)-h^{(t)}(X_t)|^2\right)^{1/2},\\
d_T&:=\max_{t\le T}\left(\E\,|h_{\mathcal S}(X^\circ)-h^{(t)}(X^\circ)|^2\right)^{1/2}.
\end{align*}
These expectations include the training observations, their replacements, and any independent algorithmic randomization. The quantity $r_T$ measures the size of the fitted weighted error at an independent observation. The quantities $s_T$ and $d_T$ measure the effect of replacing one training observation, evaluated at that original observation and at an independent observation, respectively. The distinction matters because the empirical correction is evaluated on the original training observations.

\begin{corollary}[A stability condition]
\label{cor:no_crossfit_stability}
Retain the setup, positive component variances, and target condition of Proposition~\ref{prop:no_crossfit}. Suppose the measurable fitting rule above satisfies
\[
r_T\longrightarrow0,
\qquad \sqrt T\,(s_T+d_T)\longrightarrow0.
\]
Then all conclusions of Proposition~\ref{prop:no_crossfit} hold. More precisely, with $\|U\|_2:=(\E|U|^2)^{1/2}$ for a random variable $U$,
\begin{equation}
\label{eq:stability_bounds}
\|\mathbb G_T h_{\mathcal S}\|_2
\le r_T+\sqrt T\,(s_T+2d_T),
\qquad
\E\,\mathbb P_T h_{\mathcal S}^2\le(r_T+s_T)^2.
\end{equation}
\end{corollary}

\begin{proofof}{Corollary~\ref{cor:no_crossfit_stability}}
We first replace each observation before evaluating the fitted error at that observation. This restores conditional centering term by term. The resulting terms are still dependent, so we will control their covariances by replacing two observations at a time, and then bound the change from the original fit by stability.

Replacing observation $t$ makes $h^{(t)}$ independent of $X_t$. Define
\[
U_t:=h^{(t)}(X_t)-Ph^{(t)},
\qquad
S_T:=T^{-1/2}\sum_{t=1}^T U_t.
\]
Conditional on all original observations except $O_t$, and on all replacements and algorithmic randomization, $h^{(t)}$ is fixed while $X_t$ retains its $P_X$ law. Hence $U_t$ has conditional mean zero and conditional second moment $P\{(h^{(t)})^2\}-(Ph^{(t)})^2\le P\{(h^{(t)})^2\}$. A replaced training sample has the same distribution as $\mathcal S$, so taking expectations gives $\E U_t^2\le r_T^2$.

For distinct $t,u$, the term $U_t$ still depends on $O_u$ through its fit, and $U_u$ still depends on $O_t$. To bound their covariance, replace both observations, using the same copies $O_t',O_u'$, to obtain $h^{(t,u)}$, and define
\[
V_t:=h^{(t,u)}(X_t)-Ph^{(t,u)},
\qquad
V_u:=h^{(t,u)}(X_u)-Ph^{(t,u)}.
\]
The fit $h^{(t,u)}$ uses neither $O_t$ nor $O_u$. Conditional on the doubly replaced training sample and its independent randomization, $X_t$ and $X_u$ are therefore independent with their original laws, so $\E V_tV_u=0$. The mixed products also have mean zero: if we condition on all original observations except $O_t$, and on all copies and randomization, $V_u$ is fixed and $U_t$ has conditional mean zero. Thus $\E U_tV_u=0$, and interchanging $t$ and $u$ gives $\E V_tU_u=0$. Expanding the product of differences now yields
\[
\E U_tU_u=\E\{(U_t-V_t)(U_u-V_u)\}.
\]
The difference $h^{(t)}-h^{(t,u)}$ is independent of $X_t$. Also, replacing $O_t$ preserves the independent and identically distributed law of the training sample, so $d_T$ still bounds the effect of replacing observation $u$ in that sample. Consequently,
\[
\E(U_t-V_t)^2
=\E\,\Var_P\{h^{(t)}(X)-h^{(t,u)}(X)\}
\le\E\,P\{(h^{(t)}-h^{(t,u)})^2\}
\le d_T^2.
\]
Here $\Var_P$ integrates an independent $X\sim P_X$ with the two fitted functions fixed. The same argument bounds $\E(U_u-V_u)^2$ by $d_T^2$. Cauchy--Schwarz then gives $|\E U_tU_u|\le d_T^2$. Summing the $T$ diagonal terms and the $T(T-1)$ off-diagonal terms in $\E S_T^2$ gives
\[
\E S_T^2\le r_T^2+(T-1)d_T^2.
\]

We now return to the original empirical process. Their difference is
\[
\mathbb G_T h_{\mathcal S}-S_T
=T^{-1/2}\sum_{t=1}^T\{(h_{\mathcal S}-h^{(t)})(X_t)
-P(h_{\mathcal S}-h^{(t)})\}.
\]
The first part of each summand has $L^2$ norm at most $s_T$. For the second part, Jensen's inequality gives
$\E\{P(h_{\mathcal S}-h^{(t)})\}^2\le\E\,P\{(h_{\mathcal S}-h^{(t)})^2\}\le d_T^2$.
The triangle inequality gives
\begin{align*}
\|\mathbb G_T h_{\mathcal S}-S_T\|_2
&\le T^{-1/2}\sum_{t=1}^T\Bigl(
\|(h_{\mathcal S}-h^{(t)})(X_t)\|_2
+\|P(h_{\mathcal S}-h^{(t)})\|_2\Bigr)\\
&\le\sqrt T\,(s_T+d_T).
\end{align*}
Since $\|S_T\|_2\le\{r_T^2+(T-1)d_T^2\}^{1/2}\le r_T+\sqrt T\,d_T$, adding the two bounds proves the first inequality in \eqref{eq:stability_bounds}.

It remains to control the empirical second moment used for variance estimation. Independence of $X_t$ and $h^{(t)}$, together with the unchanged law of the replaced training sample, gives $\|h^{(t)}(X_t)\|_2=r_T$. Thus
\[
\|h_{\mathcal S}(X_t)\|_2
\le\|h^{(t)}(X_t)\|_2
+\|(h_{\mathcal S}-h^{(t)})(X_t)\|_2
\le r_T+s_T.
\]
Squaring and averaging over $t$ proves the second inequality in \eqref{eq:stability_bounds}. The assumed rates make the first bound tend to zero and the expected empirical second moment tend to zero. Chebyshev's and Markov's inequalities give the two labeled-sample conditions in \eqref{eq:no_crossfit_conditions}, and Proposition~\ref{prop:no_crossfit} applies.
\end{proofof}

Entropy control and stability provide two sufficient ways to reuse all labels. Neither requires the limiting predictor to equal the regression function. In this known-representer setting the population identity is exact, so the additional restrictions address the dependence created by fitting and evaluating on the same observations.

\par\endgroup

\section{Building the predictor}
\label{app:proof_construction}

The following subsections parallel Sections~\ref{sec:prediction_classes}--\ref{sec:stacking}: prediction classes and the representer, prediction-loss fitting, variance optimization, and stacking. We give the proofs and additional conditions for each construction.

\subsection{Prediction classes and the representer}
\label{app:prediction_classes}

\begin{proofof}{Lemma~\ref{lem:index}}
By \eqref{eq:representer} with $\Delta=\gamma_0$, and since $\alpha_0(X)$ is a function of $X$,
$\theta_0=\E_P\,\alpha_0(X)\gamma_0(X)=\E_P\,\alpha_0(X)Y$. By iterated expectations given $A=\alpha_0(X)$,
$\E_P\,AY=\E_P\,A\,\bar\gamma_0(A)$. The projected regression $\bar\gamma_0\circ\alpha_0$ is square-integrable by Jensen's inequality and $\E_PY^2<\infty$. If it also belongs to $\mathcal L$, applying \eqref{eq:representer} to this function gives $\E_P\,A\,\bar\gamma_0(A)=\E_Q\,m(Z,\bar\gamma_0\circ\alpha_0)$.
\end{proofof}

In a randomized trial with constant treatment probability, a least-squares fit that includes arm intercepts has residuals that average to zero in each arm of its fitting sample. This training-sample balance does not require the held-out residual means to vanish. Remark~\ref{rem:calib_riesz} develops calibration of the representer and its balance conditions.

\begin{remark}[Calibrating the representer instead]
\label{rem:calib_riesz}
We can also seek empirical balance by calibrating the representer, following the isotonic approach of \citet{vanderlaan2024calibration}. A bounded range ensures that the optimization problem has a solution. The balance identities depend on whether the fitted values reach the range boundaries. We consider functionals whose sample evaluations depend on finitely many function values, as for means and treatment effects. The average-derivative map requires the smoothness conditions of Appendix~\ref{app:derivative_regularity} and is excluded.

Fix the samples, and let $s_1<\cdots<s_r$ collect all distinct values of $\alpha_0$ at inputs used by either term of the empirical Riesz loss. For a treatment effect, these include the counterfactual scores $\alpha_0(1,W_i)$ and $\alpha_0(0,W_i)$ in the target term. Given $v\in\R^r$, let $g_v$ be the right-continuous step function with $g_v(s_j)=v_j$, constant between successive scores and beyond the extreme scores. Assume $g_v\circ\alpha_0\in\mathcal L$ for every $v\in\R^r$. This domain condition also permits the perturbations used below.

The loss depends only on the vector $v$. Let $q_j$ be the fraction of labeled observations with $\alpha_0(X_t)=s_j$, and define the target coefficients $c_j$ by linearity so that
\[
\frac1N\sum_{i=1}^N m(Z_i,g_v\circ\alpha_0)=\sum_{j=1}^r c_jv_j,
\qquad v\in\R^r.
\]
For a finite bound $B>0$, define
\begin{equation}
\label{eq:calib_bounded_loss}
\mathcal C_B:=\{v\in\R^r:-B\le v_1\le\cdots\le v_r\le B\},
\qquad
\widehat L(v):=\sum_{j=1}^r q_jv_j^2-2\sum_{j=1}^r c_jv_j.
\end{equation}
Choose a minimizer $\widehat v\in\arg\min_{v\in\mathcal C_B}\widehat L(v)$, and set $\widehat g:=g_{\widehat v}$ and $\widehat\alpha:=\widehat g\circ\alpha_0$. The feasible set is nonempty and compact, and the loss is a continuous convex quadratic, so a minimizer exists even when some $q_j$ are zero. Differentiating the loss along the feasible segment from $\widehat v$ to any $v\in\mathcal C_B$ gives
\begin{equation}
\label{eq:calib_variational}
\sum_{j=1}^r(q_j\widehat v_j-c_j)(v_j-\widehat v_j)\ge0,
\qquad v\in\mathcal C_B.
\end{equation}

To obtain an equality, take a bounded function $h$ that vanishes at each attained range boundary: require $h(-B)=0$ if some $\widehat v_j=-B$, and $h(B)=0$ if some $\widehat v_j=B$. The perturbation $v_j^\varepsilon:=\widehat v_j+\varepsilon h(\widehat v_j)$ preserves ties. There are finitely many distinct fitted levels, so their positive gaps preserve the ordering for sufficiently small $|\varepsilon|$. The restriction on $h$ fixes the boundary levels, and the remaining levels stay inside the range. Thus $v^\varepsilon\in\mathcal C_B$ for both signs of sufficiently small $\varepsilon$. Substituting these perturbations in \eqref{eq:calib_variational} yields
\begin{equation}
\label{eq:calib_balance}
\frac1T\sum_{t=1}^{T}\widehat\alpha(X_t)h\bigl(\widehat\alpha(X_t)\bigr)
=\frac1N\sum_{i=1}^{N}m\bigl(Z_i,h\circ\widehat\alpha\bigr).
\end{equation}
Here $h\circ\widehat\alpha=g_u\circ\alpha_0$ with $u_j=h(\widehat v_j)$, so the domain condition and the definition of $c_j$ apply. If all fitted levels lie strictly inside $[-B,B]$, \eqref{eq:calib_balance} holds for every bounded $h$. At an active range boundary, general directions have only the corresponding one-sided optimality inequality.

Consequently, for every $h$ satisfying the stated boundary restriction, the estimator \eqref{eq:estimator} with $\widehat\alpha$ in place of $\alpha_0$ has the same value at $f$ and at $f+h\circ\widehat\alpha$: the change in its target average is canceled by the change in its residual correction. This is an exact empirical invariance. A larger finite $B$ need not make every fitted level interior, since a score entering only the target term can have $q_j=0$ and be pushed to a range boundary by its linear coefficient.

For an RCT with constant propensity and both labeled arms nonempty, write $q:=T_1/T\in(0,1)$ and let $g_0,g_1$ be the calibrated values in the control and treatment arms. The loss reduces to
\[
(1-q)g_0^2+qg_1^2-2(g_1-g_0),
\qquad -B\le g_0\le g_1\le B.
\]
If $B>\max\{(1-q)^{-1},q^{-1}\}$, its unique minimizer is $g_0=-(1-q)^{-1}$ and $g_1=q^{-1}$, as follows by setting the two derivatives to zero. These values satisfy the order and range constraints strictly. Hence
\[
\widehat\alpha(d,w)=\frac d q-\frac{1-d}{1-q},
\]
and \eqref{eq:calib_balance} gives invariance to arbitrary arm-specific shifts of $f$.

To distinguish this empirical property from population bias, let $\alpha$ be a fixed square-integrable working representer and $f\in\mathcal L$ be fixed. Write $\widehat\theta_\alpha(f)$ for \eqref{eq:estimator} with $\alpha$ replacing $\alpha_0$. Subtracting \eqref{eq:identity} and conditioning on $X$ gives
\[
\E\widehat\theta_\alpha(f)-\theta_0
=\E_P\{\alpha(X)-\alpha_0(X)\}\{\gamma_0(X)-f(X)\},
\]
whose absolute value is at most $\nP{\alpha-\alpha_0}\,\nP{\gamma_0-f}$ by Cauchy--Schwarz. This identity quantifies sensitivity to error in a fixed representer. The empirical balance equations alone do not establish population-bias or inference guarantees for the data-dependent $\widehat\alpha$; those require additional analysis. Calibration of the prediction itself, item~(e) of the menu in Section~\ref{sec:construction}, is studied for a mean by \citet{vanderlaan2026calibeating}.
\hfill{$\blacksquare$}
\end{remark}

\subsection{Fitting by prediction loss}
\label{app:prediction_loss}

The squared loss in \eqref{eq:prediction_loss_fit} can be replaced by a general prediction loss $\ell$. Proposition~\ref{prop:param} continues to hold under the following conditions. In addition to compactness, continuity of the predictor and penalty, the weighted predictor envelope, and $\lambda_k\toP0$, assume that $\beta\mapsto\ell(Y,f_\beta(X))$ is almost surely continuous and $\E_P\sup_{\beta\in\mathcal B}|\ell(Y,f_\beta(X))|<\infty$. The identification condition then applies to the population risk $R(\beta):=\E_P\ell(Y,f_\beta(X))$: every minimizer must represent the same function $\fstar$.

\begin{proofof}{Proposition~\ref{prop:param}}
We prove the general-loss version just stated. For squared loss, its additional conditions follow from continuity of $f_\beta$, $\E_PY^2<\infty$, and the bound $\sup_\beta\{Y-f_\beta(X)\}^2\le2Y^2+2\sup_\beta f_\beta(X)^2$, so it includes Proposition~\ref{prop:param}.

Fix $k$ and define $R_{n_k}(\beta):=n_k^{-1}\sum_{t\notin I_k}\ell(Y_t,f_\beta(X_t))$. Since $n_k\to\infty$, compactness, continuity and the integrable envelope give the uniform law of large numbers $\sup_{\beta\in\mathcal B}|R_{n_k}(\beta)-R(\beta)|\toP0$ \citep[Lemma~2.4]{NeweyMcFadden1994}, and $R$ is continuous. Continuity of $\Omega$ on the compact set $\mathcal B$ also makes the penalty bounded. Thus the bound
\[
\begin{aligned}
\sup_{\beta\in\mathcal B}|R_{n_k}(\beta)+\lambda_k\Omega(\beta)-R(\beta)|
&\le\sup_{\beta\in\mathcal B}|R_{n_k}(\beta)-R(\beta)|\\
&\quad+\lambda_k\sup_{\beta\in\mathcal B}|\Omega(\beta)|=\op(1)
\end{aligned}
\]
shows that the penalized criterion also converges uniformly, including when the penalty level is selected on the training sample.

The minimizer set of $R$ is nonempty and compact. For each $\varepsilon>0$, continuity and compactness give a positive risk gap between this set and the points of $\mathcal B$ at distance at least $\varepsilon$ from it, if any exist. Once the uniform error is less than half this gap, a minimizer of the penalized criterion cannot lie among those points. It follows that the distance from $\widehat\beta_k$ to the set of minimizers of $R$ tends to zero in probability \citep[Theorem~5.7 and its proof]{vanderVaart1998}.

We now transfer this convergence to the predictions. The map $\beta\mapsto f_\beta$ is continuous from $\mathcal B$ into the space with norm $\nPa{\cdot}$: if $\beta_j\to\beta$ then $f_{\beta_j}(X)\to f_\beta(X)$ almost surely, and $\{1+\alpha_0(X)^2\}\{f_{\beta_j}(X)-f_\beta(X)\}^2$ is dominated by four times the integrable envelope, so dominated convergence applies. On the compact set $\mathcal B$ the map is uniformly continuous, and every minimizer of $R$ is mapped to $\fstar$. Hence $\nPa{\hf_k-\fstar}\toP0$. Under the remaining conditions of Theorem~\ref{thm:cf_clt}, its inferential conclusion follows.
\end{proofof}

For bounded least-squares classes, local complexity gives a quantitative bound \citep{BartlettBousquetMendelson2005,Koltchinskii2006}. We measure the complexity of deviations within $L_2(P)$ distance $\delta>0$ of the population minimizer $\fstar$ by
\[
\mathfrak R_n(\delta):=\E\sup\left\{\frac1n\sum_{t\le n}\varepsilon_t h(X_t):h=f-\fstar,\ f\in\mathcal F,\ \nP{h}\le\delta\right\}.
\]
Here the $\varepsilon_t$ are independent Rademacher signs, independent of the data, and the expectation is over both the signs and the sample. We assume measurability of the empirical-process suprema used below.

\begin{proposition}
\label{prop:critical}
Suppose that $|Y|\le B$, that $\mathcal F\subseteq\mathcal L$ is a class of functions bounded by $B$ and star-shaped around a minimizer $\fstar$ of $f\mapsto\E_P\{Y-f(X)\}^2$ over $\mathcal F$, and that $\hf_k$ is a measurable minimizer of the empirical squared loss over $\mathcal F$ on the $n_k:=T-|I_k|$ labeled observations outside $I_k$. Let $\delta_{n}>0$ satisfy $\mathfrak R_{n}(\delta_{n})\le c\,\delta_{n}^2$, where $\mathfrak R_n(\delta)$ is the Rademacher complexity of $\{f-\fstar:f\in\mathcal F,\ \nP{f-\fstar}\le\delta\}$ for samples of size $n$. Then there is a constant $C$, depending only on $B$ and $c$, such that for each $\eta\in(0,1)$, with probability at least $1-\eta$,
\[
\nP{\hf_k-\fstar}\le C\Bigl(\delta_{n_k}+\sqrt{\log(1/\eta)/n_k}\Bigr).
\]
If moreover $|\alpha_0|\le\bar\alpha$, then $\nPa{\hf_k-\fstar}\le(1+\bar\alpha^2)^{1/2}\,\nP{\hf_k-\fstar}$.
If $n_k\to\infty$, $\delta_{n_k}\to0$, and $\E_P\alpha_0(X)^2<\infty$, then $\nPa{\hf_k-\fstar}\toP0$.
\end{proposition}

\begin{proofof}{Proposition~\ref{prop:critical}}
Fix $k$ and write $n:=n_k$, $\hf:=\hf_k$, $R(f):=\E_P\{Y-f(X)\}^2$, and $R_n(f)$ for the empirical squared loss on the $n$ training observations, whose empirical measure we denote by $P_n$; we write $Pg$ for $\E_Pg$. Let $\widehat h:=\hf-\fstar$.

\emph{Step 1: excess risk dominates squared distance.} Let $f\in\mathcal F$ and define $h:=f-\fstar$. Star-shapedness ensures that $\fstar+uh\in\mathcal F$ for $0\le u\le1$. Since $\fstar$ minimizes $R$ over $\mathcal F$, the risk along this segment is minimized at $u=0$, so its right derivative satisfies
\[
\left.\frac{d}{du}R(\fstar+uh)\right|_{u=0^+}
=-2P(Y-\fstar)h\ge0.
\]
Expanding the squared loss and using this sign gives
\begin{equation}
\label{eq:pop_lb}
R(f)-R(\fstar)=P\bigl[h^2-2(Y-\fstar)h\bigr]\ge\nP{h}^2 .
\end{equation}

\emph{Step 2: convexity localizes the minimizer.} Let $\Lambda_n(h):=R_n(\fstar+h)-R_n(\fstar)=P_n[h^2-2(Y-\fstar)h]$. By optimality, $\Lambda_n(\widehat h)\le0$. Fix $\delta>0$ and suppose that
\begin{equation}
\label{eq:boundary_gap}
\Lambda_n(h)>0\quad\text{for every }h=f-\fstar\text{ with }f\in\mathcal F\text{ and }\nP{h}=\delta .
\end{equation}
Then $\nP{\widehat h}\le\delta$. Indeed, if $\nP{\widehat h}>\delta$, define $u:=\delta/\nP{\widehat h}\in(0,1)$. By star-shapedness $\fstar+u\widehat h\in\mathcal F$, and $\nP{u\widehat h}=\delta$. By convexity of the squared loss,
$R_n(\fstar+u\widehat h)\le(1-u)R_n(\fstar)+uR_n(\hf)\le R_n(\fstar)$, so $\Lambda_n(u\widehat h)\le0$, which contradicts \eqref{eq:boundary_gap}.

\emph{Step 3: one localized deviation bound.} To establish the boundary gap, we control the difference between empirical and population excess risk on the ball
\[
\mathcal H(\delta):=\{f-\fstar:f\in\mathcal F,\ \nP{f-\fstar}\le\delta\}.
\]
For fixed $\delta>0$, define $\ell_h:=h^2-2(Y-\fstar)h$, $Z_\delta:=\sup_{h\in\mathcal H(\delta)}|(P-P_n)\ell_h|$, and $\epsilon:=\sqrt{\log(1/\eta)/n}$. Since $|h|\le2B$ and $|Y-\fstar|\le2B$, we have $|\ell_h|\le6B|h|\le12B^2$ and $\operatorname{Var}(\ell_h)\le P\ell_h^2\le36B^2\delta^2$. Conditional on the observations, each map $u\mapsto u^2-2\{Y_t-\fstar(X_t)\}u$ vanishes at zero and has derivative bounded in absolute value by $8B$ on $[-2B,2B]$. Symmetrization and contraction therefore give
\[
\E Z_\delta\le32B\mathfrak R_n(\delta).
\]
Here the absolute Rademacher supremum costs at most a factor of $2$: since $0\in\mathcal H(\delta)$, it is bounded by the sum of the two one-sided suprema, whose expectations agree by sign symmetry.

The centered losses satisfy $|\ell_h-P\ell_h|\le24B^2$. We apply Bousquet's version of Talagrand's inequality \citep{bousquet2002bennett}, in the form stated by \citet[Theorem~A.1]{BartlettBousquetMendelson2005}, to the class containing both signs of these losses. With probability at least $1-\eta$, this gives
\[
Z_\delta\le\E Z_\delta
+\sqrt{2\{36B^2\delta^2+48B^2\E Z_\delta\}}\,\epsilon
+8B^2\epsilon^2.
\]
We split the square root and use $2uv\le u^2+v^2$ to absorb the term involving $\sqrt{\E Z_\delta}\,\epsilon$ into multiples of $\E Z_\delta$ and $\epsilon^2$. Consequently, there are constants $a_1,a_2,a_3$, depending only on $B$, such that on the same event
\begin{equation}
\label{eq:local_dev}
\sup_{h\in\mathcal H(\delta)}\bigl|(P-P_n)\bigl[h^2-2(Y-\fstar)h\bigr]\bigr|\le a_1\mathfrak R_n(\delta)+a_2\,\delta\epsilon+a_3\,\epsilon^2=:\Psi(\delta).
\end{equation}
We use \eqref{eq:local_dev} only at the deterministic radius chosen below, so no peeling is needed.

\emph{Step 4: the boundary gap at the critical radius.} Take $\delta:=C(\delta_n+\epsilon)$ with $C\ge1$ to be chosen. The map $s\mapsto\mathfrak R_n(s)/s$ is nonincreasing. Indeed, let $0<s\le s'$ and $h\in\mathcal H(s')$. By star-shapedness $(s/s')h$ is again of the form $f-\fstar$ with $f\in\mathcal F$, and its norm is at most $s$, so $(s/s')h\in\mathcal H(s)$. Hence, for every realization of the signs and the data,
\[
\sup_{h\in\mathcal H(s')}\frac1n\sum_{t\le n}\varepsilon_th(X_t)=\frac{s'}{s}\sup_{h\in\mathcal H(s')}\frac1n\sum_{t\le n}\varepsilon_t\,\frac{s}{s'}h(X_t)\le\frac{s'}{s}\sup_{g\in\mathcal H(s)}\frac1n\sum_{t\le n}\varepsilon_tg(X_t),
\]
and taking expectations gives $\mathfrak R_n(s')/s'\le\mathfrak R_n(s)/s$.

Since $\delta\ge C\delta_n\ge\delta_n$, it follows that $\mathfrak R_n(\delta)\le\delta\,\mathfrak R_n(\delta_n)/\delta_n\le c\,\delta\delta_n\le c\,\delta^2/C$. Also $\epsilon\le\delta/C$. Hence
$\Psi(\delta)\le(a_1c/C+a_2/C+a_3/C^2)\,\delta^2\le\delta^2/2$ once $C$ is large enough, which fixes $C$ in terms of $B$ and $c$. On the event in \eqref{eq:local_dev}, every $h=f-\fstar$ with $f\in\mathcal F$ and $\nP{h}=\delta$ therefore satisfies, by \eqref{eq:pop_lb},
\[
\Lambda_n(h)=P\bigl[h^2-2(Y-\fstar)h\bigr]-(P-P_n)\bigl[h^2-2(Y-\fstar)h\bigr]\ge\delta^2-\Psi(\delta)\ge\delta^2/2>0 .
\]
Thus \eqref{eq:boundary_gap} holds and Step~2 gives $\nP{\widehat h}\le\delta$. If $|\alpha_0|\le\bar\alpha$, then $\nPa{\widehat h}^2=P[(1+\alpha_0^2)\widehat h^2]\le(1+\bar\alpha^2)\nP{\widehat h}^2$, which gives the stated quantitative bound for a bounded representer.

To obtain weighted consistency, suppose that $n_k\to\infty$ and $\delta_{n_k}\to0$. The probability bound gives $\nP{\widehat h}\toP0$. Write $\mathbf1\{\cdot\}$ for the indicator. Since $|\widehat h|\le2B$, truncating the squared representer at any $L>0$ gives, for every realized fit,
\[
\nPa{\widehat h}^{\,2}
\le(1+L)\nP{\widehat h}^{\,2}
+4B^2\E_P\bigl[\alpha_0(X)^2\mathbf1\{\alpha_0(X)^2>L\}\bigr].
\]
For fixed $L$, the first term converges to zero in probability. The second tends to zero as $L\to\infty$ because $\E_P\alpha_0(X)^2<\infty$. Taking these limits in order proves $\nPa{\widehat h}\toP0$.
\end{proofof}

For a deterministic sequence of classes $\mathcal F_T$ satisfying the assumptions of Proposition~\ref{prop:critical} with common constants $B$ and $c$, use their corresponding population minimizers $f_T^\star$. If the critical radii for all training folds vanish and their sizes diverge, the proposition gives $\nPa{\hf_k-f_T^\star}\toP0$. To obtain the fixed limit required by Assumption~\ref{ass:pseudo}, suppose also that $\nPa{f_T^\star-\fstar}\to0$ for some fixed $\fstar\in\mathcal L$. Then
\[
\nPa{\hf_k-\fstar}
\le\nPa{\hf_k-f_T^\star}+\nPa{f_T^\star-\fstar}\toP0.
\]
The same truncation argument shows that $L_2(P)$ convergence of the uniformly bounded population minimizers to a bounded $\fstar$ suffices. The finite summand variances in Assumption~\ref{ass:pseudo} remain required. The learning results concern empirical minimizers: applying them to data-selected regularization requires the stated penalty convergence, and applying Proposition~\ref{prop:critical} to an iterative fit requires that it attain the empirical minimum. Other fitting procedures can be used whenever their weighted consistency is established.

\subsection{Fitting for estimator variance}
\label{app:variance_construction}

We first study the residual-variance criterion used in Section~\ref{sec:variance_fitting}, then extend the analysis to the full variance at general sample-size ratios.

\begingroup
\subsubsection{Residual-variance optimization}
\label{app:variance_fit}

We first prove the guarantee in Section~\ref{sec:construction}, then give sufficient conditions for the uniform convergence of the criterion and the weighted convergence of the fitted function. Completing the square gives
\[
\E_P(R_f-c)^2=\sigma_P^2(f)+(c-\E_PR_f)^2.
\]
The same identity under $\mathbb{P}_{-k}$ shows that profiling over $c$ yields $\widehat\sigma_{P,k}^2(f)$, with $\widehat c_k=\mathbb{P}_{-k}R_{\hf_k}$. The divisor $n_k$, rather than $n_k-1$, does not affect the minimizing predictor.

\begin{proofof}{Proposition~\ref{prop:variance_fit}}
Compare $\hf_k$ with any $f\in\mathcal F$. The definition of $\varepsilon_k$ and approximate empirical optimality give
\[
\sigma_P^2(\hf_k)
\le\widehat\sigma_{P,k}^2(\hf_k)+\varepsilon_k
\le\widehat\sigma_{P,k}^2(f)+\eta_k+\varepsilon_k
\le\sigma_P^2(f)+2\varepsilon_k+\eta_k.
\]
Taking the infimum over $f$ proves \eqref{eq:variance_oracle}, even when the population infimum is not attained.

If $\varepsilon_k+\eta_k\toP0$, the bound implies $\sigma_P^2(\hf_k)\toP\inf_{f\in\mathcal F}\sigma_P^2(f)$. To identify the limiting variance, we apply the reverse triangle inequality to the $L_2(P)$ norms of the centered residuals. Since centering cannot increase this norm, Assumption~\ref{ass:pseudo} gives
\[
\bigl|\sigma_P(\hf_k)-\sigma_P(\fstar)\bigr|
\le\bigl(\E_P[\alpha_0(X)^2\{\hf_k(X)-\fstar(X)\}^2]\bigr)^{1/2}
\le\nPa{\hf_k-\fstar}=\op(1).
\]
Thus $\sigma_P^2(\fstar)=\inf_{f\in\mathcal F}\sigma_P^2(f)$. Theorem~\ref{thm:cf_clt} gives the inferential conclusions under the additional assumptions in the proposition. Finally, when $T=o(N)$, the positive finite component variances give
\[
\sigma^2(\fstar)=\frac{\sigma_P^2(\fstar)}{T}+\frac{\sigma_Q^2(\fstar)}{N}
\sim\frac1T\inf_{f\in\mathcal F}\sigma_P^2(f).
\]
\end{proofof}

Variance optimality alone need not identify a prediction function. For a mean, where $\alpha_0=1$, adding a constant to $f$ leaves $\sigma_P^2(f)$ unchanged. If a linear family has a separate, freely varying constant coefficient, fixing it leaves the attainable variances unchanged. The identification condition in the next proposition applies after any such normalization.

\begin{proposition}[Conditions for residual-variance optimization]
\label{prop:variance_fit_consistency}
Let $\mathcal F=\{f_\beta:\beta\in\mathcal B\}\subseteq\mathcal L$, and suppose that the suprema below are measurable. Suppose that $n_k\to\infty$, that $\mathcal B\subseteq\R^d$ is nonempty and compact with $d$ fixed, and that $(\beta,x)\mapsto f_\beta(x)$ is measurable and continuous in $\beta$ for $P_X$-almost every $x$. Assume
\[
\E_P\bigl[\alpha_0(X)^2Y^2\bigr]<\infty,
\qquad
\E_P\sup_{\beta\in\mathcal B}\{1+\alpha_0(X)^2\}f_\beta(X)^2<\infty.
\]
Then the minimizer set in \eqref{eq:variance_fit} is nonempty almost surely, and $\varepsilon_k\toP0$. Every measurable choice of minimizer is therefore variance-optimal in probability. If, in addition, every minimizer $\bar\beta$ of $\beta\mapsto\sigma_P^2(f_\beta)$ gives the same function $f_{\bar\beta}=\fstar$, $P$-almost surely, then every such choice satisfies $\nPa{\hf_k-\fstar}\toP0$ for each $k$.
\end{proposition}

\begin{proofof}{Proposition~\ref{prop:variance_fit_consistency}}
Fix $k$ and write $R_\beta:=R_{f_\beta}$. The functions $R_\beta$ are almost surely continuous in $\beta$, and
\[
\sup_{\beta\in\mathcal B}R_\beta^2
\le 2\alpha_0(X)^2Y^2+2\alpha_0(X)^2\sup_{\beta\in\mathcal B}f_\beta(X)^2=:H,
\qquad \E_PH<\infty.
\]
The classes $\{R_\beta\}$ and $\{R_\beta^2\}$ thus have integrable envelopes $H^{1/2}$ and $H$. We verify the uniform laws for both classes. Write $g_\beta$ for either $R_\beta$ or $R_\beta^2$. For any $\delta>0$ and each $\beta\in\mathcal B$, continuity and dominated convergence give a relative neighborhood $U_\beta$ such that
\[
\E_P\sup_{b\in U_\beta}|g_b-g_\beta|<\delta.
\]
Compactness supplies a finite cover $U_{\beta_1},\ldots,U_{\beta_J}$ of $\mathcal B$. Let $w_j:=\sup_{b\in U_{\beta_j}}|g_b-g_{\beta_j}|$. These suprema are measurable by continuity and separability. For every $b\in U_{\beta_j}$,
\[
|(\mathbb{P}_{-k}-P)g_b|
\le |(\mathbb{P}_{-k}-P)g_{\beta_j}|+\mathbb{P}_{-k}w_j+Pw_j,
\]
where $Pg:=\E_Pg$. The ordinary law of large numbers applies to the finitely many centers and envelopes, so the supremum over $b\in\mathcal B$ is at most $2\delta+\op(1)$. Since $\delta$ is arbitrary,
\[
a_k:=\sup_{\beta\in\mathcal B}|(\mathbb{P}_{-k}-P)R_\beta|\toP0,
\qquad
b_k:=\sup_{\beta\in\mathcal B}|(\mathbb{P}_{-k}-P)R_\beta^2|\toP0.
\]
The bound $\sup_\beta|PR_\beta|\le(\E_PH)^{1/2}$ now yields
\[
\varepsilon_k\le b_k+a_k\{2(\E_PH)^{1/2}+a_k\}\toP0.
\]

The empirical criterion $\beta\mapsto\widehat\sigma_{P,k}^2(f_\beta)$ is continuous almost surely and hence attains its minimum on $\mathcal B$. Choosing $c=\mathbb{P}_{-k}R_\beta$ at any minimizing $\beta$ gives a joint minimizer in \eqref{eq:variance_fit}. Proposition~\ref{prop:variance_fit}, with $\eta_k=0$, proves variance optimality.

It remains to prove convergence of the predictor under the additional identification condition. Dominated convergence makes $\beta\mapsto\sigma_P^2(f_\beta)$ continuous. It also makes $\beta\mapsto f_\beta$ continuous in $\nPa{\cdot}$, since squared differences are bounded by four times the integrable predictor envelope. The population criterion attains its minimum on the compact set $\mathcal B$. By assumption all its minimizers represent $\fstar$. For any $\epsilon>0$, the set
\[
C_\epsilon:=\{\beta\in\mathcal B:\nPa{f_\beta-\fstar}\ge\epsilon\}
\]
is compact. If it is nonempty, it contains no population minimizer, so
\[
\inf_{\beta\in C_\epsilon}\sigma_P^2(f_\beta)
-\min_{\beta\in\mathcal B}\sigma_P^2(f_\beta)>0.
\]
On the event that $2\varepsilon_k$ is smaller than this gap, \eqref{eq:variance_oracle} excludes every empirical minimizer from $C_\epsilon$. This event has probability approaching $1$. If $C_\epsilon$ is empty, every feasible predictor is already within distance $\epsilon$. Thus $\nPa{\hf_k-\fstar}\toP0$ for every measurable choice of minimizer.
\end{proofof}

The moment conditions imply $\sigma_P^2(\fstar)<\infty$. Together with $\sigma_Q^2(\fstar)<\infty$, the identification condition therefore supplies Assumption~\ref{ass:pseudo}. Under Assumptions~\ref{ass:known_riesz} and~\ref{ass:msc} and positive component variances, Proposition~\ref{prop:variance_fit} gives valid inference and the stated leading-variance optimality when $T=o(N)$. Correct specification of the regression is not required.

\subsubsection{Full-variance optimization at general sample-size ratios}
\label{app:full_variance_fit}

When labels are not asymptotically scarce, improving the labeled residuals can increase the variation of the target predictions. We can account for this tradeoff by retaining both components of $\sigma^2(f)$. To keep their weights bounded for every relative growth rate, define
\begin{equation}
\label{eq:full_normalized_risk}
\begin{split}
\rho_{N,T}&:=\frac{T}{N+T},\qquad \kappa:=\frac{NT}{N+T},\\
\mathcal V_{N,T}(f)&:=(1-\rho_{N,T})\sigma_P^2(f)
+\rho_{N,T}\sigma_Q^2(f)=\kappa\sigma^2(f).
\end{split}
\end{equation}
Thus minimizing $\mathcal V_{N,T}$ is equivalent to minimizing the full estimator variance. If $T=o(N)$, its weights approach $1$ and $0$, recovering the residual criterion of Section~\ref{sec:construction}. If $T/N$ has a positive finite limit, both weights remain positive. The formulation also covers $N=o(T)$, when the target component receives limiting weight $1$.

The centering argument used for residual-variance optimization applies to both terms. Write $M_f(Z):=m(Z,f)$ and retain $R_f:=\alpha_0(X)\{Y-f(X)\}$. Completing the square separately under $P$ and $Q$ gives
\begin{equation}
\label{eq:full_profile}
\mathcal V_{N,T}(f)
=\min_{c_P,c_Q\in\R}
\left\{(1-\rho_{N,T})\E_P(R_f-c_P)^2
+\rho_{N,T}\E_Q(M_f-c_Q)^2\right\}.
\end{equation}
For a linear family $f_\beta=q^\top\beta$, linearity of $m$ makes both summands quadratic in $\beta,c_P,c_Q$. Full-variance optimization therefore remains a least-squares problem with two centering constants, subject to the chosen constraint on $\beta$. The constants center the fitting criterion and do not enter the corrected estimator.

\paragraph{Training and evaluation in both samples.}
Estimating the $Q$-component may require using target observations to fit $f$. To preserve independence from the observations used for evaluation, we then cross-fit both samples. Fix $K\ge2$ and deterministic equal-sized partitions $I_1,\ldots,I_K$ of the labeled indices and $J_1,\ldots,J_K$ of the target indices. For clarity, suppose $K$ divides both $T$ and $N$. Otherwise, discard fewer than $K$ observations from each sample and use the resulting sample sizes; this does not change their limiting ratio. Each $\hf_k$ is measurable with respect to the observations outside $I_k$ and $J_k$, together with any external data independent of both samples.

Write $n_k:=T-|I_k|$ and $m_k:=N-|J_k|$, and define the training empirical averages
\[
\mathbb P_{-k}g:=\frac1{n_k}\sum_{t\notin I_k}g(Y_t,X_t),
\qquad
\mathbb Q_{-k}h:=\frac1{m_k}\sum_{i\notin J_k}h(Z_i).
\]
The empirical component variances and the full fitting criterion are
\begin{equation}
\label{eq:full_empirical_risk}
\begin{split}
\widehat\sigma_{P,k}^2(f)&:=\mathbb P_{-k}R_f^2-(\mathbb P_{-k}R_f)^2,\\
\widehat\sigma_{Q,k}^2(f)&:=\mathbb Q_{-k}M_f^2-(\mathbb Q_{-k}M_f)^2,\\
\widehat{\mathcal V}_k(f)&:=(1-\rho_{N,T})\widehat\sigma_{P,k}^2(f)
+\rho_{N,T}\widehat\sigma_{Q,k}^2(f).
\end{split}
\end{equation}
Profiling the empirical counterpart of \eqref{eq:full_profile} over $c_P,c_Q$ gives exactly \eqref{eq:full_empirical_risk}. The weights use the final evaluation sample sizes $T,N$. For a chosen class $\mathcal F$, we fit by minimizing $\widehat{\mathcal V}_k$ on the training observations. When a separate, independent target training sample is available, it can instead supply the $Q$-criterion, and the original estimator \eqref{eq:theta_cf} remains available. If $Q$-moments can be calculated from the labeled-data law, target observations need not enter fitting at all. The two-sample construction covers the general case.

The next bound separates the statistical error in estimating each variance from the error in optimizing their weighted sum. It does not require either a limiting sample-size ratio or a population minimizer to exist.

\begin{proposition}[Full-variance oracle bound]
\label{prop:full_variance_oracle}
Let $\mathcal F\subseteq\mathcal L$ be fixed and nonempty, with
$\E_PR_f^2+\E_QM_f^2<\infty$ for every $f\in\mathcal F$.
For each $k$, suppose that the training-measurable fit $\hf_k\in\mathcal F$ satisfies
\[
\widehat{\mathcal V}_k(\hf_k)
\le\inf_{f\in\mathcal F}\widehat{\mathcal V}_k(f)+\eta_k,
\qquad \eta_k\ge0.
\]
Assume that the quantities
\[
\varepsilon_{P,k}:=\sup_{f\in\mathcal F}
|\widehat\sigma_{P,k}^2(f)-\sigma_P^2(f)|,
\qquad
\varepsilon_{Q,k}:=\sup_{f\in\mathcal F}
|\widehat\sigma_{Q,k}^2(f)-\sigma_Q^2(f)|
\]
are measurable, and define
$\delta_k:=(1-\rho_{N,T})\varepsilon_{P,k}+\rho_{N,T}\varepsilon_{Q,k}$.
Then
\begin{equation}
\label{eq:full_variance_oracle}
0\le\mathcal V_{N,T}(\hf_k)-\inf_{f\in\mathcal F}\mathcal V_{N,T}(f)
\le2\delta_k+\eta_k.
\end{equation}
In particular, $\delta_k+\eta_k\toP0$ gives optimality in probability for the normalized full-variance criterion, without a restriction on the relative growth rates of $N$ and $T$.
\end{proposition}

\begin{proofof}{Proposition~\ref{prop:full_variance_oracle}}
The definition of $\delta_k$ gives
$\sup_{f\in\mathcal F}|\widehat{\mathcal V}_k(f)-\mathcal V_{N,T}(f)|\le\delta_k$.
For any $f\in\mathcal F$, approximate empirical optimality therefore implies
\[
\mathcal V_{N,T}(\hf_k)
\le\widehat{\mathcal V}_k(\hf_k)+\delta_k
\le\widehat{\mathcal V}_k(f)+\eta_k+\delta_k
\le\mathcal V_{N,T}(f)+\eta_k+2\delta_k.
\]
Taking the infimum over $f$ proves the bound. Its lower bound follows from $\hf_k\in\mathcal F$.
\end{proofof}

A penalty can be useful for fitting a rich class, but the proposition concerns approximate minimization of the unpenalized variance criterion. For example, a nonnegative penalty that is bounded on $\mathcal F$, multiplied by a coefficient converging to zero, contributes a vanishing amount to $\eta_k$. Selecting a nonvanishing penalty by cross-validation alone does not establish the proposition's optimality conclusion.

\paragraph{Inference after full-variance optimization.}
The fit for fold $k$ is evaluated only on $I_k$ and $J_k$. Define
\begin{equation}
\label{eq:two_sample_cf}
\hth_{\mathrm{2CF}}
:=\frac1K\sum_{k=1}^K
\left\{\frac1{|J_k|}\sum_{i\in J_k}m(Z_i,\hf_k)
+\frac1{|I_k|}\sum_{t\in I_k}\alpha_0(X_t)\{Y_t-\hf_k(X_t)\}\right\}.
\end{equation}
Let $k_P(t)$ and $k_Q(i)$ denote the folds containing labeled observation $t$ and target observation $i$, respectively. Each observation contributes one held-out summand. The variance estimator is consequently
\begin{equation}
\label{eq:two_sample_cf_variance}
\widehat V_{\mathrm{2CF}}
:=\frac1N\widehat\Var_N\!\left(m(Z_i,\hf_{k_Q(i)})\right)
+\frac1T\widehat\Var_T\!\left(\alpha_0(X_t)\{Y_t-\hf_{k_P(t)}(X_t)\}\right).
\end{equation}
Both sample variances use the usual divisor, the corresponding sample size minus one. The target variance here uses held-out predictions, whereas \eqref{eq:Vcf} uses fold-averaged predictions because its fits do not use target observations.

Full-variance optimization and valid inference require distinct conditions. Proposition~\ref{prop:full_variance_oracle} controls the achieved variance. To use the fitted function for inference, we also require the same weighted convergence and mean-square continuity as in Section~\ref{sec:finetune}, now for fits trained on both sample complements.

\begin{proposition}[Two-sample inference]
\label{prop:two_sample_full_inference}
Use the training and evaluation construction above. Suppose that Assumption~\ref{ass:known_riesz} holds, that $\hf_k\in\mathcal L$ almost surely, and that
$\E\sigma_P^2(\hf_k)+\E\sigma_Q^2(\hf_k)<\infty$ for every $k$.
Then $\E\hth_{\mathrm{2CF}}=\theta_0$.
If additionally Assumptions~\ref{ass:pseudo} and~\ref{ass:msc} hold for these fits with a deterministic limit $\fstar$ satisfying $\sigma_P^2(\fstar)>0$ and $\sigma_Q^2(\fstar)>0$, then $\hth_{\mathrm{2CF}}-\hth(\fstar)=\op\bigl(\sigma(\fstar)\bigr)$ as $N,T\to\infty$ at any relative rates.
If, in addition, $\fstar\in\mathcal F$ and the conditions of Proposition~\ref{prop:full_variance_oracle} hold with $\delta_k+\eta_k\toP0$ for every $k$, then
\begin{equation}
\label{eq:full_class_efficiency}
\frac{\sigma^2(\fstar)}{\inf_{f\in\mathcal F}\sigma^2(f)}\longrightarrow1.
\end{equation}
The ratio compares the variance at the limiting predictor with the smallest fixed-predictor variance in the class. Under the stated moment and convergence conditions, the inferential conclusions of Theorem~\ref{thm:cf_clt} hold with $(\hth_{\CF},\widehat V_{\CF})$ replaced by $(\hth_{\mathrm{2CF}},\widehat V_{\mathrm{2CF}})$.
\end{proposition}

\begin{proofof}{Proposition~\ref{prop:two_sample_full_inference}}
Let $\mathcal T_k$ be the sigma-field generated by the two training complements and the independent external data. Conditional on $\mathcal T_k$, the held-out observations in $I_k$ and $J_k$ have their original independent product laws. Lemma~\ref{lem:identity} makes the conditional mean of the expression in braces in \eqref{eq:two_sample_cf} equal to $\theta_0$. Its conditional mean squared deviation from $\theta_0$ is
\[
\frac{K}{N}\sigma_Q^2(\hf_k)+\frac{K}{T}\sigma_P^2(\hf_k).
\]
The expected variance assumptions and Tonelli's theorem make each combined fold estimate square-integrable. The tower property therefore proves exact unbiasedness after averaging over $k$.

We next compare the fitted estimator with the fixed-predictor estimator. Define $\Delta_k:=\hf_k-\fstar$ and
$\mu_k:=\E_Qm(Z,\Delta_k)=\E_P\alpha_0(X)\Delta_k(X)$, where the equality follows from the representer identity. Equal fold sizes ensure that replacing every fit by $\fstar$ in \eqref{eq:two_sample_cf} gives exactly \eqref{eq:estimator}. Hence
\[
\hth_{\mathrm{2CF}}-\hth(\fstar)=\frac1K\sum_{k=1}^K(A_k-C_k),
\]
where
\[
A_k:=\frac1{|J_k|}\sum_{i\in J_k}\{m(Z_i,\Delta_k)-\mu_k\},
\qquad
C_k:=\frac1{|I_k|}\sum_{t\in I_k}\{\alpha_0(X_t)\Delta_k(X_t)-\mu_k\}.
\]
Both terms have conditional mean zero given $\mathcal T_k$, and
\[
\E[A_k^2\mid\mathcal T_k]\le\frac{K}{N}\E_Qm(Z,\Delta_k)^2,
\qquad
\E[C_k^2\mid\mathcal T_k]\le\frac{K}{T}\nPa{\Delta_k}^2.
\]
On the training-measurable event $\{\Delta_k\in\mathcal D\}$, Assumption~\ref{ass:msc} bounds the first conditional second moment by $KM\nP{\Delta_k}^{a}/N$. The two normalized bounds converge to zero in probability by Assumption~\ref{ass:pseudo}. Conditional Chebyshev's inequality, truncated at $1$ before taking expectations, gives $A_k=\op(N^{-1/2})$ and $C_k=\op(T^{-1/2})$. For example, the probability that $\sqrt N|A_k|>\epsilon$ is at most
\[
\Pp(\Delta_k\notin\mathcal D)
+\E\min\{1,KM\nP{\Delta_k}^{a}/\epsilon^2\}\longrightarrow0.
\]
The last expectation tends to zero because its integrand is bounded and converges to zero in probability. Since $K$ is fixed, summing gives a remainder $\op(N^{-1/2}+T^{-1/2})$. Positive component variances at $\fstar$ make this $\op(\sigma(\fstar))$. Theorem~\ref{thm:clt} now gives the normal limit with normalization $\sigma(\fstar)$.

For variance consistency, the same conditional Markov argument gives
\[
\frac1N\sum_{i=1}^N m(Z_i,\Delta_{k_Q(i)})^2=\op(1),
\qquad
\frac1T\sum_{t=1}^T\alpha_0(X_t)^2\Delta_{k_P(t)}(X_t)^2=\op(1).
\]
Indeed, the conditional expectations of the corresponding fold averages are bounded by $M\nP{\Delta_k}^{a}$ and $\nPa{\Delta_k}^2$ whenever $\Delta_k\in\mathcal D$. The sample standard deviation is a seminorm, so the difference between each fitted and fixed-predictor sample standard deviation is bounded by the sample standard deviation of their difference, which is $\op(1)$ by these displays. The ordinary laws of large numbers for the fixed-predictor summands then yield component variance consistency. Their positively weighted sum is ratio-consistent for $\sigma^2(\fstar)$ at every relative sample-size rate. Slutsky's lemma proves studentization.

Finally, the standard-deviation triangle inequality gives
\[
|\sigma_P(\hf_k)-\sigma_P(\fstar)|\le\nPa{\Delta_k},
\qquad
|\sigma_Q(\hf_k)-\sigma_Q(\fstar)|\le\{\E_Qm(Z,\Delta_k)^2\}^{1/2}.
\]
Both bounds are $\op(1)$ under the stated stability conditions. Consequently
$\mathcal V_{N,T}(\hf_k)-\mathcal V_{N,T}(\fstar)=\op(1)$, since the two weights lie in $[0,1]$. Combining this with \eqref{eq:full_variance_oracle}, and using $\fstar\in\mathcal F$, gives
\[
0\le\mathcal V_{N,T}(\fstar)-\inf_{f\in\mathcal F}\mathcal V_{N,T}(f)\longrightarrow0.
\]
The expression is deterministic, so convergence in probability is ordinary convergence. Moreover, $\mathcal V_{N,T}(\fstar)$ is bounded below by
$\min\{\sigma_P^2(\fstar),\sigma_Q^2(\fstar)\}>0$.
Its difference from the class infimum tends to zero, so the infimum is eventually bounded away from zero as well. Dividing and canceling the common factor $\kappa$ proves \eqref{eq:full_class_efficiency}.
\end{proofof}

The next corollary gives primitive conditions under which fitting the full criterion produces the stability just used. A limiting sample-size ratio is needed here to identify one deterministic limiting predictor; it was not needed for the oracle bound or the inference result conditional on stability.

\begin{corollary}[Compact prediction classes]
\label{cor:full_variance_compact}
Let $\mathcal F=\{f_\beta:\beta\in\mathcal B\}$ satisfy the compactness, measurability, continuity, and $P$-moment conditions of Proposition~\ref{prop:variance_fit_consistency}. In addition, suppose that $(\beta,z)\mapsto m(z,f_\beta)$ is measurable and continuous in $\beta$ for $Q$-almost every $z$, and that
\[
\E_Q\sup_{\beta\in\mathcal B}m(Z,f_\beta)^2<\infty.
\]
Let $N,T\to\infty$, and assume that $\rho_{N,T}\to\rho_0\in[0,1]$.
Then the empirical full-variance criterion attains its minimum almost surely, and $\varepsilon_{P,k},\varepsilon_{Q,k}\toP0$ for each $k$.
Suppose further that every minimizer of
\[
\beta\longmapsto(1-\rho_0)\sigma_P^2(f_\beta)+\rho_0\sigma_Q^2(f_\beta)
\]
gives the same function $\fstar$, $P$-almost surely. Every measurable approximate empirical minimizer with $\eta_k\toP0$ then satisfies $\nPa{\hf_k-\fstar}\toP0$. If Assumption~\ref{ass:known_riesz} holds, the mean-square continuity inequality in Assumption~\ref{ass:msc} holds on $\mathcal F-\mathcal F$, and both component variances at $\fstar$ are positive, all conclusions of Proposition~\ref{prop:two_sample_full_inference} follow.
\end{corollary}

\begin{proofof}{Corollary~\ref{cor:full_variance_compact}}
Proposition~\ref{prop:variance_fit_consistency} supplies the uniform law for the $P$-component. The same finite-cover argument applies under $Q$ to the classes $\{m(Z,f_\beta)\}$ and $\{m(Z,f_\beta)^2\}$: their continuity and integrable envelopes give uniform laws for the first and second moments, hence $\varepsilon_{Q,k}\toP0$. The empirical variances are continuous in $\beta$ almost surely, so compactness gives an empirical minimizer.

To compare the empirical criterion with the limiting criterion, define
\[
e_k:=\sup_{\beta\in\mathcal B}
\left|\widehat{\mathcal V}_k(f_\beta)
-\{(1-\rho_0)\sigma_P^2(f_\beta)+\rho_0\sigma_Q^2(f_\beta)\}\right|.
\]
The uniform component errors and the changing weight give
\[
e_k\le\delta_k+|\rho_{N,T}-\rho_0|
\sup_{\beta\in\mathcal B}\{\sigma_P^2(f_\beta)+\sigma_Q^2(f_\beta)\}\toP0,
\]
because $\delta_k\toP0$ and the moment envelopes bound both population variances uniformly over $\mathcal B$.

Dominated convergence makes the limiting criterion continuous and the map $\beta\mapsto f_\beta$ continuous in $\nPa{\cdot}$. For each $\epsilon>0$, the compact set
$\{\beta\in\mathcal B:\nPa{f_\beta-\fstar}\ge\epsilon\}$ either is empty or is separated from the minimum of the limiting criterion by a positive gap, because it contains no minimizer. Comparing an approximate empirical minimizer with a population minimizer bounds its excess limiting criterion by $2e_k+\eta_k$. Once this bound is smaller than the gap, the fitted parameter cannot lie in that set. Since $2e_k+\eta_k\toP0$, this proves weighted convergence of the fitted function. Finally, the same envelopes bound $\E\sigma_P^2(\hf_k)+\E\sigma_Q^2(\hf_k)$ uniformly for fits in $\mathcal F$. The remaining conditions of Proposition~\ref{prop:two_sample_full_inference} are assumed explicitly, so its conclusions apply.
\end{proofof}

\begin{remark}[Optimal prediction for a mean]
\label{rem:full_variance_mean}
For a mean with $Q=P_X$ and $\alpha_0=1$, write $\rho=\rho_{N,T}$. Conditioning on $X$ separates the residual variance into two terms:
\[
\Var_P\{Y-f(X)\}
=\E_P\Var_P(Y\mid X)+\Var_P\{\gamma_0(X)-f(X)\}.
\]
Substituting this identity into \eqref{eq:full_normalized_risk} and completing the square gives
\[
\mathcal V_{N,T}(f)
=(1-\rho)\E_P\Var_P(Y\mid X)
+\rho(1-\rho)\Var_P\gamma_0(X)
+\Var_P\{f(X)-(1-\rho)\gamma_0(X)\}.
\]
Over all square-integrable predictors, the minimizers are therefore
$f=(1-\rho)\gamma_0+c$, with $c\in\R$. When $T=N$, the optimal adjustment is $\gamma_0/2$, up to a constant; when $T=o(N)$, its coefficient tends to $1$. This is the unrestricted-function counterpart of the scalar calculation in Remark~\ref{rem:variance_targets}. It shows why regression fitting and full-variance optimization can have different targets even with correct specification.
\hfill{$\blacksquare$}
\end{remark}

\begin{remark}[Prediction loss and estimator variance]
\label{rem:variance_targets}
For a fixed scalar prediction $p$, assume that the moments below are finite and the denominators are positive. Least-squares regression of $Y$ on $p(X)$ with a free intercept has slope $\lambda_{\mathrm{LS}}:=\Cov_P(Y,p(X))/\Var_P(p(X))$. Expanding $T\sigma^2(\lambda p)$ as a quadratic in $\lambda$ shows that the minimizer of the full variance $\sigma^2(\lambda p)$ is
\[
\lambda_{\mathrm{var}}
:=\frac{\Cov_P\bigl(\alpha_0(X)Y,\alpha_0(X)p(X)\bigr)}
{\Var_P\bigl(\alpha_0(X)p(X)\bigr)+(T/N)\Var_Q\bigl(m(Z,p)\bigr)}.
\]
For a mean with $Q=P_X$ and $\alpha_0=1$, this becomes $\lambda_{\mathrm{var}}=\lambda_{\mathrm{LS}}/(1+T/N)$. PPI++ estimates the variance-optimal scalar by a plug-in rule \citep[Section~6, Example~6.1]{angelopoulos2024ppiplusplus}; it already targets inferential precision within this scalar family. The formula above extends that principle to the known-representer setting and underlies the comparator in Appendix~\ref{app:procedure}. Our direct fitting rule \eqref{eq:variance_fit} minimizes $\sigma_P^2(f)$ over $\mathcal F$, with leading-variance optimality when $T=o(N)$. The full-variance criterion \eqref{eq:full_normalized_risk} retains both variance components.
\hfill{$\blacksquare$}
\end{remark}

\par\endgroup

\subsection{Stacking and selection}
\label{app:stacking}

\emph{Stacking} combines fitted predictors using weights chosen to minimize the estimated full variance. Let $\hf_{j,k}$, for $j=1,\dots,J$, be $J$ learners cross-fitted on common folds $I_1,\dots,I_K$, with fold weights $\omega_k:=|I_k|/T$, and let $\hth_{\CF,j}$ be their estimates. For weights $v\in\R^J$ with $\sum_jv_j=1$, we use the combined predictor $\sum_jv_j\hf_{j,k}$.

To estimate the variance of the resulting estimator, let $\widehat S_Q$ be the sample covariance matrix of the vectors $\bigl(\sum_k\omega_k m(Z_i,\hf_{j,k})\bigr)_{j\le J}$ over the target observations, and let $\widehat S_P$ be that of $\bigl(\alpha_0(X_t)\{Y_t-\hf_{j,k(t)}(X_t)\}\bigr)_{j\le J}$ over the labeled observations. Define $\widehat\Sigma:=\widehat S_Q/N+\widehat S_P/T$. Linearity of $m$ and $\sum_jv_j=1$ give
\begin{equation}
\label{eq:stack}
\hth_{\CF}(v)=\sum_{j=1}^{J}v_j\,\hth_{\CF,j},
\qquad
\widehat V_{\CF}(v)=v^{\top}\widehat\Sigma\,v.
\end{equation}
Given a set $\mathcal V\subseteq\{v\in\R^J:\sum_jv_j=1\}$, we choose $\widehat v\in\arg\min_{v\in\mathcal V}v^{\top}\widehat\Sigma v$, a convex quadratic program when $\mathcal V$ is the simplex. The weights depend on the labels, so exact unbiasedness need not hold. Consistency of the weights makes their estimation negligible at the standard-error scale because the component estimators share the same target. The corollary's limiting criterion comes from multiplying the variance by $NT/(N+T)$.

\begin{corollary}[Stacking]
\label{cor:stack}
Let $J$ be fixed and $N,T\to\infty$. Suppose that Assumption~\ref{ass:known_riesz} holds and that each learner $j\le J$ satisfies Assumptions~\ref{ass:pseudo} and~\ref{ass:msc} with limit $\fstar_j$. Let $S_Q$ and $S_P$ be the covariance matrices of $\bigl(m(Z,\fstar_j)\bigr)_{j\le J}$ under $Q$ and of $\bigl(\alpha_0(X)\{Y-\fstar_j(X)\}\bigr)_{j\le J}$ under $P$. Suppose that $T/(N+T)\to\rho\in[0,1]$, that $\mathcal V$ is a nonempty closed convex subset of $\{v\in\R^J:\sum_jv_j=1\}$, and that $q(v):=v^{\top}\{\rho S_Q+(1-\rho)S_P\}v$ has a unique minimizer $v^\star$ over $\mathcal V$, with $q(v^\star)>0$. Then $\widehat v\toP v^\star$ and $\{\hth_{\CF}(\widehat v)-\theta_0\}/\widehat V_{\CF}(\widehat v)^{1/2}\toD\mathcal N(0,1)$.
\end{corollary}

The scalar adjustment discussed in Remark~\ref{rem:variance_targets} is the case of learners $p$ and $0$ with affine weights $(\lambda,1-\lambda)$.

The empirical minimizer in Corollary~\ref{cor:stack} may be set-valued. We interpret $\widehat v\toP v^\star$ to mean that, for every $\varepsilon>0$, with probability approaching $1$ the minimizer set is nonempty and all its elements lie within distance $\varepsilon$ of $v^\star$. When the set is empty, we assign a fixed value $v_0\in\mathcal V$. Statements involving a selected weight vector apply to any measurable selection. The proof below establishes this convergence, including nonemptiness with probability approaching $1$.

\begin{proofof}{Corollary~\ref{cor:stack}}
Let $\kappa:=NT/(N+T)$, so that $1/\kappa=1/N+1/T$, and let $\rho_{N,T}:=T/(N+T)$. Then $\kappa\widehat\Sigma=\rho_{N,T}\widehat S_Q+(1-\rho_{N,T})\widehat S_P$. For each learner, the variance-estimation argument in the proof of Theorem~\ref{thm:cf_clt} shows that the empirical second moment of the difference between its fitted and limiting summands is $\op(1)$ in each sample. Cauchy--Schwarz and the law of large numbers therefore give entrywise convergence $\widehat S_Q\toP S_Q$ and $\widehat S_P\toP S_P$, without requiring positive component variances. Since $J$ is fixed,
\[
\|\kappa\widehat\Sigma-\bar S\|_{\mathrm{op}}\toP0,
\qquad \bar S:=\rho S_Q+(1-\rho)S_P.
\]
Consequently, the convex criterion $\widehat q(v):=v^{\top}\kappa\widehat\Sigma v$ converges to $q(v)=v^{\top}\bar Sv$ uniformly on every bounded set: for each finite $R$,
\[
\sup_{\|v\|\le R}|\widehat q(v)-q(v)|
\le R^2\|\kappa\widehat\Sigma-\bar S\|_{\mathrm{op}}=\op(1).
\]

We now localize every empirical minimizer, allowing $v^\star$ to lie on the boundary of $\mathcal V$. Fix $\varepsilon>0$ and let $C_\varepsilon:=\{v\in\mathcal V:\|v-v^\star\|=\varepsilon\}$. If $C_\varepsilon$ is nonempty, compactness, continuity, and uniqueness of $v^\star$ give
\[
\eta_\varepsilon:=\min_{v\in C_\varepsilon}\{q(v)-q(v^\star)\}>0.
\]
Uniform convergence on this set and convergence at $v^\star$ imply that, with probability approaching $1$, $\widehat q(v)>\widehat q(v^\star)$ for every $v\in C_\varepsilon$. On this event, no feasible $u$ with $\|u-v^\star\|>\varepsilon$ can satisfy $\widehat q(u)\le\widehat q(v^\star)$: the feasible segment from $v^\star$ to $u$ would cross $C_\varepsilon$, and convexity of $\widehat q$ would give a point there with value at most $\widehat q(v^\star)$.

The continuous function $\widehat q$ attains its minimum on the nonempty compact set $\mathcal V\cap\{v:\|v-v^\star\|\le\varepsilon\}$. The preceding comparison shows that this is a global minimum and that every global minimizer lies strictly within distance $\varepsilon$ of $v^\star$. If $C_\varepsilon$ is empty, convexity implies that all of $\mathcal V$ lies within that distance, so $\mathcal V$ itself is compact and the same conclusions hold. This proves $\widehat v\toP v^\star$ under the convention above. Matrix convergence then gives
\[
\kappa\,\widehat V_{\CF}(\widehat v)=\widehat v^{\top}\kappa\widehat\Sigma\widehat v
\toP v^{\star\top}\bar Sv^\star>0.
\]

For each $j$, Lemma~\ref{lem:quasi_oracle} and Assumption~\ref{ass:pseudo} give $\hth_{\CF,j}-\hth(\fstar_j)=\op(\kappa^{-1/2})$, and $\Var\,\hth(\fstar_j)=\sigma^2(\fstar_j)=O(1/\kappa)$, so $\hth_{\CF,j}-\theta_0=\Op(\kappa^{-1/2})$. Since $\sum_j\widehat v_j=\sum_jv^\star_j=1$,
\[
\hth_{\CF}(\widehat v)-\hth_{\CF}(v^\star)=\sum_{j=1}^{J}(\widehat v_j-v^\star_j)(\hth_{\CF,j}-\theta_0)=\op(1)\,\Op(\kappa^{-1/2})=\op(\kappa^{-1/2}).
\]
To obtain the normal limit, define $\fstar_{v^\star}:=\sum_jv^\star_j\fstar_j$, which belongs to the linear space $\mathcal L$. Since \eqref{eq:estimator} is affine in $f$ and the weights sum to $1$,
\[
\hth_{\CF}(v^\star)-\hth(\fstar_{v^\star})
=\sum_jv^\star_j\{\hth_{\CF,j}-\hth(\fstar_j)\}
=\op(\kappa^{-1/2}).
\]
The variance of this fixed-predictor estimator satisfies
\[
\kappa\,\sigma^2(\fstar_{v^\star})
=v^{\star\top}\{\rho_{N,T}S_Q+(1-\rho_{N,T})S_P\}v^\star
\to v^{\star\top}\bar Sv^\star>0.
\]
Thus both the fitting remainder and the effect of estimating the weights are negligible relative to $\sigma(\fstar_{v^\star})$. The preceding convergence of $\kappa\,\widehat V_{\CF}(\widehat v)$ also gives $\widehat V_{\CF}(\widehat v)/\sigma^2(\fstar_{v^\star})\toP1$. Theorem~\ref{thm:clt} applies to the fixed function $\fstar_{v^\star}$, and these two comparisons allow Slutsky's lemma to give the asserted studentized normal limit.
\end{proofof}

Greedy selection admits a direct argument over the finite menu. Under the same assumptions on the learners and sample sizes, suppose that $j^\star$ uniquely minimizes $\bar S_{jj}$ and that $\bar S_{j^\star j^\star}>0$. Choose $\widehat j\in\arg\min_{j\le J}\widehat\Sigma_{jj}$. If $J=1$ the selection is immediate. Otherwise, the gap $\min_{j\ne j^\star}(\bar S_{jj}-\bar S_{j^\star j^\star})$ is positive, and matrix convergence gives
\[
\max_{j\le J}|\kappa\widehat\Sigma_{jj}-\bar S_{jj}|\toP0.
\]
Hence $\Pp(\widehat j=j^\star)\to1$. The fixed-weight argument above applies to learner $j^\star$, and the selected estimate and variance equal those of that learner with probability approaching $1$. Its studentized normal limit therefore also holds for greedy selection.

\clearpage
\section{Empirical materials and additional results}
\label{app:empirical}

This appendix documents the outcomes, AI inputs, learning methods, and simulation procedure used in the empirical application, followed by the detailed results.

\subsection{Outcomes and their construction}
\label{app:outcomes}

Participants were adults aged 21--40 whose household income in 2019 did not exceed 300\% of the federal poverty level. They lived in counties of northern Illinois and north-central Texas. Enrollment closed in October 2020, and payments ran from November 2020 to October 2023. Outcomes come from interviewer-administered surveys at baseline, midline (after about 18 months of payments) and endline (after about 30 months), supplemented by monthly online surveys \citep{vivalt2024,bartik2024}. Table~\ref{tab:outcomes} lists the fourteen outcomes, their analysis cohort sizes, and the full-label benchmark effects $\Delta_n$.

\begin{table}[ht]
\centering
\caption{Outcomes and full-label benchmark treatment effects (difference in means with all labels observed). Individual income is significant at the 10\% level only and is retained as the earnings component of household income.}
\label{tab:outcomes}
\small
\begin{tabular}{@{}lrrrrr@{}}
\toprule
Outcome & $n$ & Control mean & Effect & SE & $t$ \\
\midrule
\multicolumn{6}{@{}l}{\emph{Expenditure} (monthly \$)} \\
Total expenditure & 2,991 & 3976.19 & $+$329.73 & 68.77 & $+$4.79 \\
Food and beverage & 2,971 & 967.18 & $+$61.07 & 19.69 & $+$3.10 \\
Rent & 2,960 & 628.53 & $+$65.34 & 25.40 & $+$2.57 \\
Car payment and insurance & 2,985 & 368.54 & $+$28.04 & 14.18 & $+$1.98 \\
\multicolumn{6}{@{}l}{\emph{Employment and income}} \\
Weekly work hours & 2,926 & 30.19 & $-$1.784 & 0.791 & $-$2.26 \\
Employment status (share) & 2,932 & 0.7475 & $-$0.0298 & 0.0151 & $-$1.97 \\
Participant and partner hours & 2,841 & 40.79 & $-$2.308 & 1.003 & $-$2.30 \\
Labor supply index (SD units) & 2,926 & 0.0000 & $-$0.0852 & 0.0384 & $-$2.22 \\
Household income (\$k per year) & 2,855 & 47.39 & $-$4.826 & 1.269 & $-$3.80 \\
Individual income (\$k per year) & 2,860 & 35.38 & $-$2.153 & 1.158 & $-$1.86 \\
\multicolumn{6}{@{}l}{\emph{Financial health and saving}} \\
Household savings share & 2,726 & 0.2053 & $+$0.0435 & 0.0125 & $+$3.48 \\
Individual savings share & 2,789 & 0.2137 & $+$0.0410 & 0.0127 & $+$3.22 \\
Additive financial-health index (0--40) & 2,885 & 18.19 & $+$0.6992 & 0.2879 & $+$2.43 \\
Self-assessed financial health (SD units) & 2,885 & 2.4437 & $+$0.0950 & 0.0387 & $+$2.46 \\
\bottomrule
\end{tabular}
\end{table}

To obtain a single post-treatment value per participant for each outcome, we pool survey waves using the original studies' rules. Special codes are treated as missing, and a valid ``no'' is coded as zero.

\paragraph{Employment and income.} We pool weekly work hours, employment status, participant and partner hours, household income and individual income from the midline and endline surveys as $Y_i=0.30\,Y_i^{\mathrm{mid}}+0.70\,Y_i^{\mathrm{end}}$. If a participant answered only one wave, we replace the missing wave by its mean within the participant's treatment arm. Incomes are in thousands of dollars per year, exclude the transfer, and are not winsorized. To form the labor supply index, we standardize pooled hours and pooled employment using their respective control-arm means and standard deviations, then average the standardized values.

\paragraph{Expenditure.} Spending is reported in repeated online surveys and in the midline and endline interviews, with different recall periods. Weekly online amounts are converted to monthly amounts, and a matched ratio puts overlapping interview items on the online scale. Within each source the periods are pooled, with weights 20\%, 30\% and 50\% for the three online study years and 30\% and 70\% for midline and endline, and a missing period is replaced by the participant's mean over observed periods. Each source is winsorized from above at its 99th percentile, and the two are combined with weights 30\% (online) and 70\% (interview), renormalized when one source is missing. Components are then summed into food, rent, car and total expenditure.

\paragraph{Financial health and saving.} Each item is pooled using the expenditure rules. The additive financial-health index sums ten financial well-being items, each scored 0--4, and is defined only when all ten are answered. The self-assessed component is this index divided by its standard deviation in the control group. The household savings share is total household financial savings divided by annual household income, and the individual share is individually held savings divided by annual individual income, both capped at one. The data release gives no formula for the two shares. These definitions reproduce the published control-group mean and standard deviation.

\subsection{Prompts and AI-generated features}
\label{app:prompts}

The features of class M4 come from three prompts: a benchmark without theory, a theory-guided prompt, and a prompt with two household personas. The theory-guided prompt asks how liquidity constraints affect spending and how the payment affects work, with the budget identity linking spending and saving to earnings and payments. The personas contrast smoothing a temporary payment with living paycheck to paycheck. All three prompts use a profile built from baseline responses only. For each participant, we query them separately at \$1,000 and at \$50, withholding the treatment label, outcome labels and identifiers. They request JSON describing averages over months 12--36. Their core text follows.

{\footnotesize
\begin{verbatim}
A. Benchmark, no theory
SYSTEM  You simulate how a specific adult in the United States actually behaves.
USER    {PROFILE}
        Starting next month this person receives ${AMOUNT} per month for 36 months,
        unconditionally, and knows the payments stop after month 36.
        Return only JSON with their average over months 12-36:
        {"monthly_spending": $, "monthly_household_earnings": $, "liquid_savings": $}

B. Theory-guided
SYSTEM  You are an economist predicting one household's behavior with a simple model.
USER    {PROFILE}
        Starting next month this person receives ${AMOUNT} per month for 36 months,
        unconditionally, and knows the payments stop after month 36.
        Use two considerations:
        1. Liquidity. Households with little savings, high-cost debt or no way to
           cover a $400 emergency tend to use money as it arrives; others can spread
           a temporary payment over many years. How constrained is this household?
        2. Work. More money can change how much household members work, in either
           direction (less need to work; or better job search, transport, childcare).
           How would this household's earnings change?
        Then make sure spending + saving = household earnings + payments.
        Return only JSON with averages over months 12-36:
        {"p_constrained": 0-1, "monthly_spending": $,
         "monthly_household_earnings": $, "monthly_saving": $}

C. Two-agent mixture
SYSTEM  You are a real person with ordinary limits on time and willpower. {TYPE}
        TYPE = SMOOTHER: "You have some savings and spread any temporary windfall
               over the coming years."
        TYPE = HAND-TO-MOUTH: "You live paycheck to paycheck and use money in the
               month it arrives."
USER    {PROFILE} + the same payment text and JSON as Prompt A.
\end{verbatim}
}

We adapt these prompts to each outcome by adding the study context and guardrails (for example, ``do not assume the payment raises or lowers any quantity you are asked for'') and the outcome's definition and output field. For prompt~C, we make one call per persona. We use the same model and settings as for the predictions $p$. Every numeric field becomes one feature, giving each participant feature vectors $q(1,w)$ and $q(0,w)$.

Two additional prompts do not depend on the payment and are asked once: a liquidity classification, following the hand-to-mouth definition of \citet{kaplan2014model}, and a classification for household income (hand-to-mouth type, adjustable hours, binding work constraint). The employment prompt returns, along with the prediction, scores between 0 and 1 for four mechanisms (precautionary, investment, present focus and obligation). We retain a feature if it is answered for at least 80\% of participants under each payment. We fill missing values with the median and add a missing indicator when more than 1\% are missing.

Only M4 uses these features. In the linear predictor $f(d,w)=b_d+c\,p(d,w)+\beta^{\top}h(d,w)$, they enter as columns of $h$, taking values $q(d,w)$ for both labeled and target draws within arm $d$. We cross-fit the learners that use these inputs. We do not impose the structural formulas of the economic model, and the probability of being constrained is simply another column. Table~\ref{tab:features} lists the features used for each reported M4 value.

\begin{table}[ht]
\centering
\caption{AI-generated features in the reported M4 values.}
\label{tab:features}
\footnotesize
\begin{tabular}{@{}lp{0.5\linewidth}l@{}}
\toprule
Outcome & Features & Theory content \\
\midrule
Total expenditure & in-fold top 10 of 22 columns from prompts A, B, C, liquidity, zero-shot $p$ & B, C, liquidity \\
Household savings & blend of zero- and few-shot $p$; B's probability of constraint & B \\
Individual savings & B's probability of constraint & B \\
Individual income & B's individual-income field & B \\
Household income & A and B fields; income classification & B, classification \\
Employment & fields of two employment prompts & four mechanisms \\
Labor index & employment-prompt fields; count of other jobs & four mechanisms \\
Food & prompt A's food field; zero-shot $p$ & none \\
Car & prompt A's three fields and its vehicle-expense field & none \\
Financial-health index & prompt A fields & none \\
Financial self-rating & prompt A's liquid savings & none \\
Weekly hours & count of other jobs & none \\
Rent, couple hours & fields of the main few-shot answer & none \\
\bottomrule
\end{tabular}
\end{table}

\subsection{The learner menu and its inputs}
\label{app:learners}

Each class uses the learner menu in Table~\ref{tab:learners} with the inputs specified below. M1 uses three learners and M2--M4 use five. All learners fit arm intercepts. Within arm $d$, labeled and target predictions use the inputs $(d,p(d,w),h(d,w))$. Learner settings are common across outcomes, apart from the specified input blocks and data-selected ridge penalty.

\begin{table}[ht]
\centering
\caption{The common learner menu. $H$ denotes the class's inputs, with the prediction $p$ in the first column.}
\label{tab:learners}
\small
\begin{tabular}{@{}lp{0.74\linewidth}@{}}
\toprule
Learner & Fit \\
\midrule
Linear & Least squares of $Y$ on $H$ after removing arm means. \\
Shrunken slopes & The same fit, with every slope $\widehat\beta_j$ except that of $p$ multiplied by $\max(0,1-\widehat s_j^{\,2}/\widehat\beta_j^{\,2})$, where $\widehat s_j^{\,2}$ is its estimated sampling variance. \\
Ridge & Ridge regression with $p$ unpenalized. Five-fold validation within each outer training fold selects a penalty from 33 logarithmically spaced values between $10^{-2}$ and $10^6$, using the fitting criterion. Columns are scaled within the current training subset; penalized effective degrees of freedom are capped at half its size. \\
Isotonic & A monotone map of $p$ plus a linear function of the other columns, by backfitting. \\
Boosted trees & Histogram gradient boosting on $(H,D)$ with up to 100 updates, depth 3 and step factor 0.05. Each tree update is centered within arm, profiling out the arm intercepts. \\
\bottomrule
\end{tabular}
\end{table}

Class M1 uses the zero-shot prediction alone, with the linear, isotonic and boosted-tree learners. Class M2 adds the outcome's baseline covariates, M3 uses the few-shot prediction with the same covariates, and M4 adds the features of Appendix~\ref{app:prompts}. The covariates form a block specified in advance for each outcome (two to six columns; 24 for participant and partner hours; 94 for the financial-health index), with the exceptions listed in Table~\ref{tab:inputs}.

For total spending in M4, the ten feature columns are those most associated with the residual of a random forest on the 177-column core, ranked again inside each outer and inner training fold. The resampling experiment conditions on the source records and cached input blocks, including their prior construction and selection. Preprocessing during fitting, including imputation and ridge scaling, uses the current training subset.

\begin{table}[ht]
\centering
\caption{Cells with a wider or narrower input set.}
\label{tab:inputs}
\footnotesize
\begin{tabular}{@{}lll@{}}
\toprule
Outcome & Class & Inputs \\
\midrule
Total spending & M4 & $p$, 177-column core, the top 10 feature columns ranked in every fit \\
Food & M4 & $p$, zero-shot $p$, prompt A's food field, 99 covariates \\
Car & M4 & $p$, prompt A's three fields, its vehicle-expense field, insurance covariate \\
Own income & M4 & $p$, 2 covariates, prompt B's income field \\
Household saving & M4 & average of few- and zero-shot $p$, probability of constraint \\
Own saving & M2, M3 & $p$, an 800-column covariate pool \\
Own saving & M4 & $p$, 2 covariates, probability of constraint \\
Financial self-rating & M4 & $p$, its 6 covariates, prompt A's liquid savings \\
\bottomrule
\end{tabular}
\end{table}

\subsection{The simulation procedure}
\label{app:procedure}

\paragraph{Predictions.} Before resampling, we submit the zero-shot and few-shot prompts for each outcome and participant in the analysis cohort, once with each payment. Few-shot worked examples match the payment named in the prompt. Participants used as worked examples are excluded from the cohorts of M3 and M4.

\paragraph{Step 1: independent draws.} Let $n_d$ be the number of source records in arm $d\in\{0,1\}$. Each replicate draws $T_d:=\mathrm{round}(0.1n_d)$ labeled records and $N_d:=n_d-T_d$ target records independently with replacement from that arm. The four samples are independent conditional on the source records and cached inputs. Draw positions can repeat source records, and labeled and target samples can overlap. The outcome is observed only for labeled positions. The benchmark $\Delta_n$ is the difference between the two empirical arm outcome means.

The same replicate seed gives matched draws across methods within each analysis cohort and matched draws and folds across fitting criteria. We partition labeled draw positions into $K=10$ folds separately within each arm, with sizes differing by at most 1. The folds split independent draw positions, so copies of the same source record can enter different folds. Denote the labeled and target positions in arm $d$ by $L_d$ and $U_d$, and its labeled positions in fold $k$ by $I_{d,k}$. The fold weight for that arm is $\omega_{d,k}:=|I_{d,k}|/T_d$.

\paragraph{Step 2: corrected arm means.} We fit $\hf_k$ on labeled positions outside fold $k$, pooling the two arms and using their respective intercepts. Target draws do not enter these fits. Free arm constants cancel from each corrected arm mean. To prevent arbitrary arm intercepts from affecting the variance estimate, we normalize each fold's predictor to zero at the same fixed source-covariate record in each arm, without using its outcome; $\hf_k$ below denotes this normalized predictor. Define the out-of-fold residuals $e_t:=Y_t-\hf_{k(t)}(d,W_t)$ for $t\in L_d$ and averaged target predictions $\bar f_{di}:=\sum_k\omega_{d,k}\hf_k(d,W_i)$ for $i\in U_d$. The estimate is
\begin{equation}
\label{eq:empirical_armwise}
\widehat\mu_d:=\frac{1}{N_d}\sum_{i\in U_d}\bar f_{di}
  +\frac{1}{T_d}\sum_{t\in L_d}e_t,
\qquad
\hth_{\CF}:=\widehat\mu_1-\widehat\mu_0.
\end{equation}
This applies the mean-estimation case of \eqref{eq:theta_cf} within each arm, where the representer is $1$. For fixed predictors, independent labeled and target sampling gives
\[
V(f)=\sum_{d=0}^1\left\{
\frac{\Var_d\{Y-f(d,W)\}}{T_d}
+\frac{\Var_d\{f(d,W)\}}{N_d}\right\},
\]
where $\Var_d$ is variance under the empirical distribution of source records in arm $d$. For fitted predictors, the plug-in estimate is $\widehat V_{\CF}:=\sum_d\{s_{L_d}^2(e)/T_d+s_{U_d}^2(\bar f_d)/N_d\}$, with each sample variance using its sample size minus 1 as denominator. The cross-fitting argument of Section~\ref{sec:finetune} applies to these armwise means under the corresponding convergence and moment conditions; exact unbiasedness alone does not establish interval coverage.

\paragraph{Step 3: combining learners.} Each learner in the class produces an estimate by Step 2. Replacing the four scalar sample variances by sample covariance matrices across learners, while retaining their sample-size divisors, gives $\widehat\Sigma$. Stacking minimizes $v^{\top}\widehat\Sigma v$ over the simplex, greedy selection chooses the learner with the smallest diagonal entry, and equal weighting averages the learners. Each rule uses the same cross-fitted residuals and target predictions, and its interval uses variance $v^{\top}\widehat\Sigma v$. Stacking weights are estimated from the replicate's data; their inferential justification is asymptotic as in Appendix~\ref{app:stacking}.

\paragraph{Step 4: fixed PPI and PPI++.} Fixed PPI uses the saved zero-shot prediction $p(d,w)$ with coefficient 1 in each arm. PPI++ estimates separate unrestricted coefficients,
\[
\widehat\lambda_d:=
\frac{\widehat\Cov_{L_d}\{Y,p(d,W)\}}
{(1+T_d/N_d)\,\widehat\Var_{L_d\cup U_d}\{p(d,W)\}},
\]
and uses $\widehat\lambda_d p(d,w)$ in the corrected arm means. The covariance uses labeled draws, and the prediction variance pools labeled and target draws within arm, with sample denominators. A constant prediction receives coefficient zero. These unrestricted coefficients use the full samples within each arm rather than cross-fitting, so PPI++ relies on the usual asymptotic plug-in justification.

\paragraph{Step 5: summaries.} Over 500 replicates we compute the empirical standard deviation of each estimator, its variance reduction relative to the labeled-only difference in means on the same cohort, and the fraction of intervals $\hth\pm1.96\,\widehat V^{1/2}$ containing $\Delta_n$. Empirical standard errors and variance reductions use variation across replicates; coverage uses the plug-in variance. The plug-in variance retains both labeled and target variance components, whatever fitting criterion is used.

\subsection{Additional results}
\label{app:full}

\begin{figure}[ht]
\centering
\includegraphics[width=\linewidth]{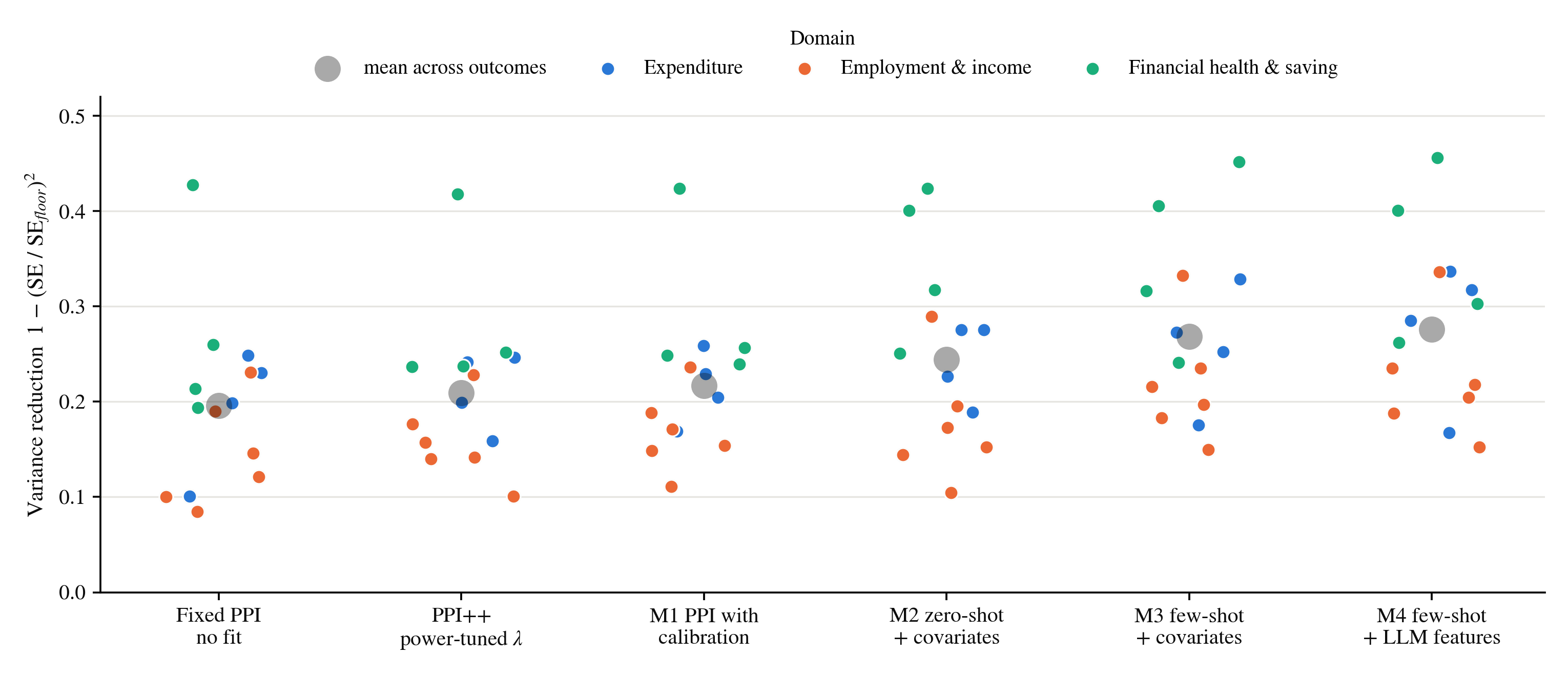}
\caption{Variance reduction by method, with 10\% of each arm labeled. Fixed PPI and PPI++ use the zero-shot prediction; M1--M4 are cross-fitted and stacked. Small points are outcomes, colored by domain and jittered; large gray points are means over the fourteen outcomes.}
\label{fig:ladder}
\end{figure}

Tables~\ref{tab:rq1} and~\ref{tab:class_summary_full} give the detailed 10\% results behind Figure~\ref{fig:heatmap} and Table~\ref{tab:class_summary} in the main text. Table~\ref{tab:objective} compares fitting criteria. At the same 10\% label budget, Figure~\ref{fig:ladder} summarizes the classes, and Table~\ref{tab:stack} compares learner-combination rules. The cohorts for classes M3 and M4 exclude the participants used as few-shot examples.

\begin{table}[ht]
\centering
\caption{Variance reduction $r$, empirical standard error and coverage of the 95\% interval for fixed PPI, PPI++ and the best cross-fitted stacked class (of M1--M4), on the same 500 replicates, with 10\% of each arm labeled. Blocks: expenditure (monthly \$); employment and income; financial health and saving. Each $r$ is relative to the labeled-only floor of its own cohort. When the best class is M3 or M4 the cohort excludes the participants used as few-shot examples, so its SE is not comparable with the Floor column. The best class is identified separately for each outcome from the reported results.}
\label{tab:rq1}
\footnotesize
\setlength{\tabcolsep}{4.2pt}
\begin{tabular}{@{}l r rrr rrr rrr@{}}
\toprule
 & Floor & \multicolumn{3}{c}{Fixed PPI} & \multicolumn{3}{c}{PPI++} & \multicolumn{3}{c}{Best fine-tuned class} \\
\cmidrule(lr){2-2}\cmidrule(lr){3-5}\cmidrule(lr){6-8}\cmidrule(l){9-11}
Outcome & SE & $r$ & SE & cov. & $r$ & SE & cov. & $r$ & SE & cov. \\
\midrule
Total spending & 223 & 0.199 & 200 & 0.938 & 0.199 & 200 & 0.934 & 0.337 & 175 & 0.948 \\
Food & 61.2 & 0.230 & 53.7 & 0.958 & 0.241 & 53.3 & 0.952 & 0.285 & 52.5 & 0.942 \\
Rent & 89.5 & 0.248 & 77.6 & 0.944 & 0.246 & 77.7 & 0.940 & 0.329 & 73.3 & 0.938 \\
Car & 42.6 & 0.101 & 40.4 & 0.960 & 0.159 & 39.0 & 0.964 & 0.189 & 38.3 & 0.956 \\
\addlinespace[2pt]
Work hours & 2.36 & 0.085 & 2.26 & 0.968 & 0.142 & 2.19 & 0.974 & 0.235 & 2.28 & 0.970 \\
Employed & 0.0477 & 0.121 & 0.0447 & 0.938 & 0.101 & 0.0452 & 0.936 & 0.205 & 0.0430 & 0.938 \\
Couple hours & 3.18 & 0.100 & 3.02 & 0.942 & 0.140 & 2.95 & 0.946 & 0.173 & 2.89 & 0.946 \\
Labor index & 0.115 & 0.146 & 0.107 & 0.960 & 0.157 & 0.106 & 0.964 & 0.218 & 0.118 & 0.940 \\
Household income & 3.89 & 0.231 & 3.41 & 0.952 & 0.228 & 3.42 & 0.948 & 0.336 & 3.17 & 0.964 \\
Own income & 3.67 & 0.190 & 3.30 & 0.960 & 0.177 & 3.33 & 0.958 & 0.195 & 3.29 & 0.962 \\
\addlinespace[2pt]
Household saving & 0.0394 & 0.194 & 0.0354 & 0.936 & 0.237 & 0.0344 & 0.934 & 0.262 & 0.0338 & 0.938 \\
Own saving & 0.0428 & 0.214 & 0.0380 & 0.942 & 0.237 & 0.0374 & 0.942 & 0.318 & 0.0354 & 0.936 \\
Financial-health index & 0.930 & 0.427 & 0.703 & 0.954 & 0.418 & 0.709 & 0.954 & 0.456 & 0.685 & 0.948 \\
Financial self-rating & 0.122 & 0.260 & 0.105 & 0.942 & 0.252 & 0.106 & 0.940 & 0.406 & 0.0943 & 0.952 \\
\midrule
Mean over outcomes & & 0.196 & & 0.950 & 0.210 & & 0.949 & 0.282 & & 0.948 \\
\bottomrule
\end{tabular}
\end{table}

\begin{table}[ht]
\centering
\caption{Variance reduction $r$ and coverage of the 95\% interval across fourteen outcomes, with 10\% labels per arm. M1--M4 use cross-fitted stacking.}
\label{tab:class_summary_full}
\small
\begin{tabular}{@{}lrrrrr@{}}
\toprule
Method & Mean & Median & Min & Max & Mean coverage \\
\midrule
Fixed PPI, $f=p$ & 0.196 & 0.196 & 0.085 & 0.427 & 0.950 \\
PPI++ & 0.210 & 0.213 & 0.101 & 0.418 & 0.949 \\
M1: calibrated zero-shot $p$ & 0.217 & 0.217 & 0.111 & 0.424 & 0.948 \\
M2: zero-shot $p$ and covariates & 0.244 & 0.239 & 0.104 & 0.424 & 0.948 \\
M3: few-shot $p$ and covariates & 0.268 & 0.247 & 0.150 & 0.452 & 0.947 \\
M4: few-shot $p$, covariates and AI features & 0.276 & 0.273 & 0.153 & 0.456 & 0.947 \\
\bottomrule
\end{tabular}
\end{table}

At 10\%, taking the best reported class for each outcome gives a descriptive mean variance reduction of 28.2\%, compared with 27.6\% for M4 used for every outcome. The intervals for these classes contain the benchmark in 93.6--97.0\% of replicates, with a mean of 94.8\%.

Table~\ref{tab:stack} compares stacking with the other combination rules and with each learner used alone. Across the 42 outcome--class cells in M2--M4, ridge has mean $r$ only 0.003 below stacking. The shortfall is 0.047 for boosted trees and ranges from 0.126 to 0.187 for shrunken slopes, linear regression, and isotonic regression. The latter learners lose precision on the widest input sets: in M4, the linear and isotonic fits have mean variance reductions of -0.053 and -0.117, indicating greater variance than the labeled-only estimator. Greedy selection falls short of stacking by 0.009 and equal weighting by 0.043 in mean variance reduction. Equal weighting retains weight on the learners that lose precision. Coverage is at least 0.92 for every rule in every reported cell.

\begin{table}[ht]
\centering
\caption{Combining the members against using one alone, with 10\% of each arm labeled. First row: mean cross-fitted $r$
of the stacked learner. Other rows: mean difference in $r$ from stacking on the same replicates; negative is lower.
The M2--M4 column averages the 42 cells in which all five members exist; M1's input is the prediction
alone, so it has no ridge or shrunken-slope member (--). Last column: lowest 95\% coverage over those cells.}
\label{tab:stack}
\footnotesize
\begin{tabular}{@{}lrrrrrr@{}}
\toprule
 & M1 & M2 & M3 & M4 & M2--M4 & Min.\ cov. \\
\midrule
Stacking: mean $r$ & 0.217 & 0.244 & 0.268 & 0.276 & 0.263 & 0.924 \\
\midrule
\multicolumn{7}{@{}l}{\emph{Other combination rules}} \\
Greedy: the member with the smallest $\widehat V_{\CF}$ & $-$0.009 & $-$0.008 & $-$0.009 & $-$0.010 & $-$0.009 & 0.924 \\
Equal weights & $-$0.002 & $-$0.015 & $-$0.015 & $-$0.100 & $-$0.043 & 0.924 \\
\midrule
\multicolumn{7}{@{}l}{\emph{Single members, used alone}} \\
Ridge, $p$ unpenalized & -- & $-$0.004 & $-$0.001 & $-$0.003 & $-$0.003 & 0.928 \\
Boosted trees & $-$0.041 & $-$0.046 & $-$0.047 & $-$0.048 & $-$0.047 & 0.928 \\
Shrunken slopes & -- & $-$0.046 & $-$0.046 & $-$0.286 & $-$0.126 & 0.930 \\
Linear & $-$0.003 & $-$0.067 & $-$0.068 & $-$0.329 & $-$0.155 & 0.926 \\
Isotonic in $p$ & $-$0.016 & $-$0.084 & $-$0.083 & $-$0.393 & $-$0.187 & 0.930 \\
\bottomrule
\end{tabular}
\end{table}

\paragraph{Fitting for prediction and variance.}
To compare fitting for prediction with fitting for inferential precision, we fit the same learner menu under three criteria on matched draws and folds. Prediction loss is the pooled mean squared error on the current training sample. To define the variance criteria, let $s_d^2$ denote a sample variance on that sample within arm $d$. The residual and full criteria are
\[
J_R(f):=\sum_{d=0}^1\frac{s_d^2\{Y-f(d,W)\}}{T_d},
\qquad
J_V(f):=J_R(f)+\sum_{d=0}^1\frac{s_d^2\{f(d,W)\}}{N_d}.
\]
The denominators are the final labeled and target sample sizes. The prediction variances in $J_V$ use training covariates, whose within-arm distribution matches that of the target covariates. Thus full-variance fitting accounts for target-sample variation without using the target evaluation draws.

For a training sample with $m_d$ rows in arm $d$, profiling free arm intercepts reduces minimization of $J_R$ to weighted least squares, with row weights proportional to \mbox{$1/\{T_d(m_d-1)\}$}. Completing the square for full-variance fitting multiplies these weights by $1+T_d/N_d$ and divides the within-arm centered response by the same factor. Linear and ridge fits solve the resulting quadratic problems, and isotonic backfitting and centered tree updates use the same transformed response and weights. Slope shrinkage and finite iterative fits remain regularized or approximate procedures. Ridge selects its penalty by five-fold inner validation of the criterion used for fitting.

Table~\ref{tab:objective} compares the three fitting criteria, with stacking applied to each fitted menu. Mean variance reduction across the 56 outcome--class cells is 25.1\% under prediction loss, 24.5\% under residual variance, and 24.9\% under full variance. Thus all three criteria yield similar precision, with a slight average advantage for prediction loss. \rev{Residual-variance fitting does slightly worse because it ignores the target-sample variance, which is not negligible at a 10\% label share. Full-variance fitting includes it and does about as well as prediction loss.} Mean coverage is 94.7--94.8\% under all three criteria.

\begin{table}[t!]
\centering
\caption{Mean variance reduction $r$ of the stacked learner over fourteen outcomes, using the same 500 replicates and 10\% labels per arm. Differences use unrounded results; negative values indicate less variance reduction than prediction-loss fitting. Mean coverage averages all 56 outcome--class cells.}
\label{tab:objective}
\small
\begin{tabular}{@{}lrrrrr@{}}
\toprule
Fitting criterion & M1 & M2 & M3 & M4 & Mean coverage \\
\midrule
Prediction loss & 0.217 & 0.244 & 0.268 & 0.276 & 0.948 \\
Residual variance & 0.212 & 0.238 & 0.261 & 0.267 & 0.947 \\
Full variance & 0.217 & 0.242 & 0.265 & 0.271 & 0.947 \\
\midrule
Residual $-$ loss & $-$0.0049 & $-$0.0065 & $-$0.0069 & $-$0.0088 & $-$0.0006 \\
Full $-$ loss & $+$0.0002 & $-$0.0017 & $-$0.0031 & $-$0.0047 & $-$0.0008 \\
\bottomrule
\end{tabular}
\end{table}

\end{document}